\documentclass[a4paper,12pt]{article}

\usepackage{diagrams}
\usepackage[top=1.0in,right=1.0in,left=1.0in,bottom=1.0in]{geometry}
\usepackage{enumerate}
\usepackage{psfrag}
\usepackage{hyperref}

\usepackage[dvips]{graphicx}
\usepackage[usenames]{color}
\usepackage{psfrag}

\usepackage{amsmath}
\usepackage{amssymb}
\usepackage{amsthm}

\theoremstyle{plain}
\newtheorem{theorem}{Theorem}[section]
\newtheorem{lemma}[theorem]{Lemma}

\theoremstyle{definition}
\newtheorem{defn}[theorem]{Definition}

\newcommand\ignore[1]{}

\newcommand{\droptext}[2]
{
\begin{array}{c}
\\[#1]
#2
\end{array}
}

\newcommand{\lowqedhere}[1]
{
\eqno{\droptext{#1}{\qedhere}}
}

\newcommand\jskip{4pt}

\newcommand\pL{\mathrm{L}}
\newcommand\pR{\mathrm{R}}

\newcommand{\jbeginproof}{\begin{proof}\vspace{-\jskip}}
\newcommand{\jbeginenumerate}{\begin{enumerate}\vspace{-\jskip}}
\newcommand{\jnoindent}{\vspace{-\jskip}\noindent}

\newcommand{\End}{\ensuremath{\mathrm{End}}}

\newcommand{\Hom}{\mathrm{Hom}}

\newcommand{\cat}[1]{\ensuremath{\mathbf{#1}}}
\newcommand{\id}{\ensuremath{ \mathrm{id} }}

\newcommand\C{\ensuremath{\mathrm{C}}}

\newcommand{\pb}{\;\!}
\newcommand{\ma}{\pb\!}

\newcommand\nfrac
{
\begin{smallmatrix}
1
\\[1.5pt]
\hline
\\[-1.6pt]
n
\end{smallmatrix}
}

\newcommand\jbigl{\raisebox{-0.30pt}{\textrm{\large (}\hspace{-0.3pt}}}
\newcommand\jbigr{\hspace{-1.0pt}\mbox{\raisebox{-0.30pt}{\textrm{\large )}}}}

\renewcommand{\dag}{\ensuremath{\dagger}}

\newcommand{\proofnote}[1]{&&\hspace{-20pt}\textrm{\emph{\small #1}}}
\renewcommand{\to}{\mbox{\begin{diagram}[width=4pt] {} & \rTo \end{diagram}}\ma}

\newcommand{\into}{\mbox{\begin{diagram}[width=4pt] {}&\rInto\end{diagram}}\ma}

\newarrow{Double}{|}{=}{=}{=}{=}

\newarrow{mapsto}{|}{-}{-}{-}{>}
\renewcommand{\mapsto}{\mbox{\begin{diagram}[width=4pt] {}&\rMapsto\end{diagram}}\ma}

\newcommand{\lessa}{\hspace{-15pt}}

\newcommand{\vc}[1]
{
\begin{array}{c}
#1
\end{array}
}

\begin{document}

\title{
  Categorical formulation of\\ finite-dimensional quantum algebras
}

\author{Jamie Vicary
\\
Oxford University Computing Laboratory
\\
\texttt{jamie.vicary@comlab.ox.ac.uk}
}

\date{June 22, 2010}
\maketitle

\begin{abstract}
We describe how \dag\-Frobenius monoids give the correct categorical description of certain kinds of finite-dimensional `quantum algebras'. We develop the concept of an \emph{involution monoid}, and use it to construct a correspondence between finite-dimensional C*\-algebras and certain types of \dag\-Frobenius monoids in the category of Hilbert spaces. Using this technology, we recast the spectral theorems for commutative C*\-algebras and for normal operators into an explicitly categorical language, and we examine the case that the results of measurements do not form finite sets, but rather objects in a finite Boolean topos. We describe the relevance of these results for topological quantum field theory.
\end{abstract}

\section{Introduction}

The main purpose of this paper is to describe how \dag\-Frobenius monoids are the correct tool for formulating various kinds of finite-dimensional `quantum algebras'. Since \dag\-Frobenius monoids have entirely geometrical axioms, this gives a new way to look at these traditionally algebraic objects.

This difference in perspective can be thought of as moving from an `internal' to an `external' viewpoint. Traditionally, we formulate a C*\-algebra as the set of elements of a vector space, along with extra structure that specifies how to multiply elements, find a unit element, apply an involution and take norms. This is an `internal' view, since we are dealing directly with the elements of the set. The `external' alternative is to `zoom out' in perspective: we can no longer discern the individual elements of the C*\-algebra, but we can see more clearly how it relates to other vector spaces, and these relationships give an alternative way to completely define the C*\-algebra. This metaphor is made completely precise by category theory, and the passage between these two types of viewpoint is familiar in categorical approaches to algebra.

We proceed in Section~\ref{strucdagcat} by introducing our categorical setting, monoidal \dag\-categories with duals, and defining an involution monoid, a categorical axiomatization of an involutive algebra. Section~\ref{frobmonsec} introduces \dag\-Frobenius monoids, and explores some useful properties of them. We specialize to the category of Hilbert spaces in Section~\ref{hilbsect}, and make the connection between \dag\-Frobenius monoids and finite-dimensional C*\-algebras precise.

An important aspect of the conventional study of C*\-algebras are the spectral theorems, for commutative C*\-algebras and for normal operators. The \dag\-Frobenius perspective on C*\-algebras allows these theorems to be presented categorically in the finite-dimensional case, and we explore this in Section~\ref{genspec}. We also use the \dag\-Frobenius monoid formalism to explore the construction of alternative quantum theories.

This work is relevant to the study of two-dimensional open-closed topological quantum field theories (TQFTs), which model the quantum dynamics of string-like topological structures which can merge together and split apart. It was shown by Lauda and Pfeiffer \cite{lp05-ocs} that such a theory is defined by a symmetric Frobenius monoid equipped with extra structure. If we also add the physical requirement that the theory should be \emph{unitary} \cite{b06-qq} then these become symmetric \dag\-Frobenius monoids, and thus finite-dimensional C*\-algebras by Lemma~\ref{unitarylemma} and the results of Section~\ref{hilbsect}. These are precisely the correct kinds of algebras with which to construct a state-sum triangulation model for the TQFT~\cite{fhk92-ltft,lp06-ssc}, and so we can deduce the following: the two-dimensional open-closed TQFTs which arise from a state sum on a triangulation are precisely the \emph{unitary} such TQFTs, up to multiplication by a scalar factor.

\ignore{
Our results are also relevant for two-dimensional conformal field theories. It is known that such theories can be modelled in a similar way to two-dimensional topological field theories \cite{frs02-tft1, ffrs05-tcf}, except that rather than constructing the Frobenius monoids in \cat{Hilb}, the category of Hilbert spaces and bounded linear maps, they are constructed in a general modular tensor category. Suppose that our modular tensor category is the category of modules for some bialgebra, and is equipped with a monoidal forgetful functor to \cat{Hilb}, sending each module to its underlying vector space. Then any \dag\-Frobenius monoid in our modular tensor category descends to a \dag\-Frobenius monoid in \cat{Hilb}, which will be equivalent to a finite-dimensional C*-algebra as described here, and a special Frobenius algebra in the modular tensor category can then be directly obtained which is of the form required for the construction of a conformal field theory.
}

The results presented here are closely tied to finite-dimensional algebras. The author is aware of some work in progress on infinite-dimensional generalizations \cite{ah10-csid}, which requires significant changes to the underlying algebraic structures. However, the importance of the finite-dimensional case should not be underestimated. In the study of topological quantum field theory, in particular, it is often necessary to restrict to finite-dimensional algebras for the constructions to be well-defined, as a consequence of compactness of the topological category.

The construction described here can be generalized far beyond the scope of the current paper. In future work, we will describe how higher-dimensional `quantum algebras' can be described as \dag\-Frobenius pseudoalgebras, `weakened' forms of Frobenius algebras which live in a monoidal 2-category. This extends results of Day, McCrudden and Street~\cite{dms03-da, s04-fmp}. These higher-dimensional quantum algebras include the fusion C*\-categories of considerable importance in the representation theory of quantum groups~\cite{k95-qg} and in topological quantum field theory~\cite{bk01-ltc}.

\subsubsection*{Why \dag\-Frobenius monoids?}

The key property of \dag\-Frobenius monoids which makes them so useful is contained in the following observation, due to Coecke, Pavlovic and the author \cite{cpv08-dfb}. Let $(V,m,u)$ be an associative, unital algebra on a complex vector space $V$, with multiplication map $m : V \otimes V \to V$ and unit map $u : \mathbb{C} \to V$. We can map any element $\alpha \in V$ into the algebra of operators on $V$ by constructing its right action, a linear map $R _\alpha := m \circ (\id _A \otimes \alpha) : V \to V$. We draw this right action in the following way:
\[
\psfrag{p}{\hspace{0pt}$\alpha$}
\includegraphics[scale=1]{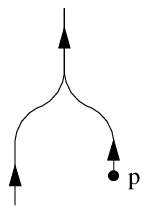}
\]
The diagram is read from bottom to top. This is a direct representation of our definition of $R _\alpha$: vertical lines represent the vector space $V$, the dot represents preparation of the state $\alpha$, and the merging of the two lines represents the multiplication operation $m: V \otimes V \to V$. If $V$ is in fact a Hilbert space we can then construct the adjoint map \mbox{$R _\alpha {}^\dag : V \to V$}. Will this adjoint also be the right action of some element of $V$?

In the case that $(V,m,u)$ is in fact a \emph{\dag\-Frobenius monoid}, the answer is yes. We draw the adjoint $R _\alpha {}^\dag$ by flipping the diagram on a horizonal axis, but keeping the arrows pointing in their original direction:
\[
\psfrag{d}{$\alpha ^\dag$}
\includegraphics{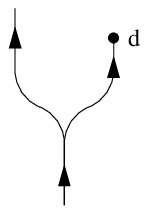}
\]
The splitting of the line into two represents the adjoint to the multiplication, and the dot represents the linear map $\alpha ^\dag : V \to \mathbb{C}$. The multiplication and unit morphisms of the \mbox{\dag\-Frobenius} monoid, along with their adjoints, must obey the following equations (see Definition~\ref{defdagfrob}):
\[
\vc{\includegraphics{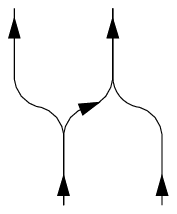}}
=
\vc{\includegraphics{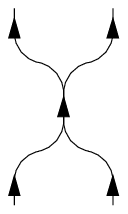}}
=
\vc{\includegraphics{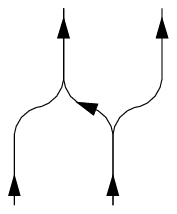}}
\hspace{40pt}
\vc{\includegraphics{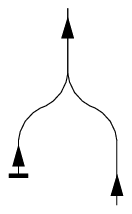}}
=
\vc{\includegraphics{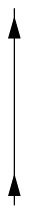}}
=
\vc{\includegraphics{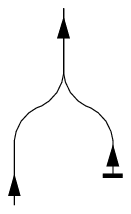}}
\]
On the left are the Frobenius equations, and on the right are the unit equations. The short horizontal bar in the unit equations represents the unit for the monoid, and the straight vertical line represents the identity homomorphism on the monoid. In fact, we also have two extra equations, since we can take the adjoint of the unit equations. We can use a unit equation and a Frobenius equation to redraw the graphical representation of $R _\alpha {}^\dag$ in the following way:
\[
{
\hspace{-10pt}
\psfrag{d}{$\alpha ^\dag$}
\vc{\includegraphics{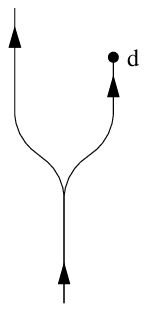}}
\hspace{-10pt}
}
=
{
\hspace{-10pt}
\psfrag{d}{$\alpha ^\dag$}
\vc{\includegraphics{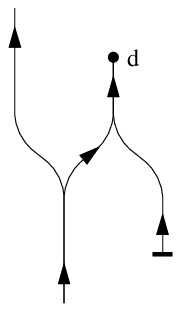}}
\hspace{-10pt}
}
=
{
\hspace{-10pt}
\psfrag{p}{$\alpha ^\dag$}
\vc{\includegraphics{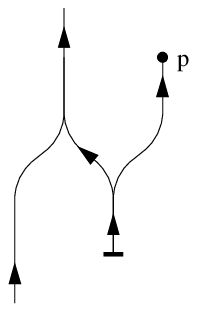}}
\hspace{-10pt}
}
=
{
\hspace{-10pt}
\psfrag{s}{$\alpha ^\dag$}
\vc{\includegraphics{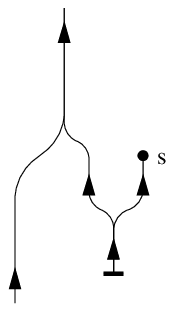}}
\hspace{-10pt}
}
\]
We therefore see that the adjoint of $R _\alpha$ is indeed a right-action of some element: $R _{\alpha} {}^\dag = R _{\alpha'}$, for $\alpha ' = (\id _{A} \otimes \alpha ^\dag) \circ m^\dag \circ u$.

To better understand this transformation $\alpha {\mapsto} \alpha'$    we apply it twice to evaluate $(\alpha')'$, using the Frobenius and unit equations and the fact that the \dag\-functor is an involution:
\[
\vspace{-10pt}
\psfrag{a}{\hspace{11pt}$(\alpha')'$}
{
\hspace{-20pt}
\vc{\includegraphics{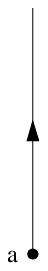}}
\hspace{-0pt}
}
=
{
\hspace{-5pt}
\psfrag{a}{\hspace{-0pt}$(\alpha' ) ^\dag$}
\vc{\includegraphics{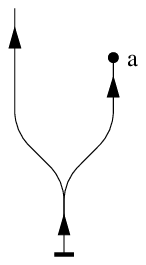}}
\hspace{-0pt}
}
=
{
\hspace{-0pt}
\psfrag{a}{\hspace{-0pt}$(\alpha ^\dag) ^\dag$}
\vc{\includegraphics{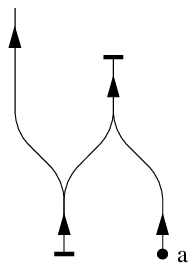}}
\hspace{-0pt}
}
=
{
\hspace{-0pt}
\psfrag{a}{\hspace{11pt}$\alpha$}
\vc{\includegraphics{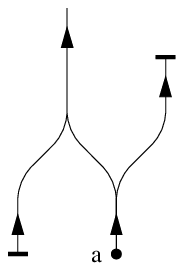}}
\hspace{-0pt}
}
=
\psfrag{a}{\hspace{11pt}$\alpha$}
{
\hspace{-0pt}
\vc{\includegraphics{graphics/newalphainv-3}}
\hspace{-15pt}
}
\]
We see that $(\alpha')'=\alpha$, and so the operation $\alpha \mapsto \alpha'$ is an involution. Since taking the adjoint $R_\alpha \mapsto R_\alpha ^\dag$ is also clearly an involution, the mapping of elements of the monoid into the ring of operators on $V$ is therefore \emph{involution-preserving}, as it maps one involution into another. We shall see that the mapping is injective and preserves the multiplication and unit of $(V,m,u)$, so in fact we have a fully-fledged involution-preserving monoid embedding as described by Lemmas \ref{embedmonoidlemma} and \ref{involembedding}.

This observation is one reason why \dag\-Frobenius monoids are such powerful tools. In fact, given that the algebra of operators on $V$ is a C*\-algebra with $*$-involution given by operator adjoint, and since any involution-closed subalgebra of a C*\-algebra is also a C*\-algebra, we have already shown that every \dag\-Frobenius monoid in \cat{Hilb} can be given a C*\-algebra norm.

\subsubsection*{Overview of paper}

We begin with a description of the categorical structure that we will use to express our results. The categories we will be working with are \emph{monoidal \dag\-categories with duals}, with nontrivial coherence requirements between the monoidal structure, \dag\-structure and duality structure. These can be seen as not-necessarily-symmetric versions of the strongly compact-closed categories of Abramsky and Coecke \cite{ac04-csqp, ac05-apt}.

We then describe the concept of an \emph{involution monoid}, a categorical version of the traditional concept of a $*$-algebra, which replaces the antilinear involution with a linear `involution' from an object to its dual. We prove some general results on involution monoids, \mbox{\dag\-Frobenius} monoids and the relationships between them, and give a definition of a \emph{special unitary \dag\-Frobenius monoid}. In \cat{Hilb}, the category of finite-dimensional complex Hilbert spaces and continuous linear maps, these monoids have particularly good properties, which we explore. We then use these properties to demonstrate in Theorem~\ref{maincalgthm} that special unitary \dag\-Frobenius monoids in \cat{Hilb} are the same as finite-dimensional \mbox{C*\-algebras}.

The spectral theorem for finite-dimensional commutative C*\-algebras is an important classical result, and we develop a way to express it using the \dag\-Frobenius toolkit. We first summarize a result from \cite{cpv08-dfb}, that the category of commutative \mbox{\dag\-Frobenius} monoids in \cat{Hilb} is equivalent to the opposite of \cat{FinSet}, the category of finite sets and functions. We generalize this by defining a monoidal \dag\-category to be \emph{spectral} if its category of commutative \dag\-Frobenius monoids is a finitary topos. We also consider the spectral theorem for normal operators, and give a way to phrase it in an abstract categorical way using the concept of \emph{internal diagonalization}.

Nontrivial examples of spectral categories are provided by categories of unitary representations of finite groupoids $\cat{Hilb ^G}$, where \cat{G} a finite groupoid. In such a category, the spectrum of a commutative generalized C*\-algebra
--- that is, the spectrum of a commutative \mbox{\dag\-Frobenius} monoid internal to the category --- is not a set, but an object in a finitary Boolean topos \cat{FinSet ^G}. Categories of the form \cat{Hilb^G} can therefore be thought of as providing alternative settings for quantum theory, in which the logic of measurement outcomes --- while still Boolean --- has a richer structure. On a technical level, we also note that this gives a new way to extract a finite groupoid from its representation category, as it is well-known that the groupoid \cat{G} can be identified in \cat{FinSet ^G} as the smallest full generating subcategory.

\subsubsection*{Acknowledgements}

The comments of the anonymous referees have greatly improved this article, and I am also grateful to Samson Abramsky, Bruce Bartlett, Bob Coecke, Chris Heunen, Chris Isham and Dusko Pavlovic for useful discussions. I am also grateful for financial support from EPSRC and QNET. Commutative diagrams are rendered using Paul Taylor's diagrams package.

\section{Structures in \dag\-categories}
\label{strucdagcat}

\subsubsection*{The \dag\-functor}
Of all the categorical structures that we will make use of, the most fundamental is the \mbox{\emph{\dag\-functor}}. It is an axiomatization of the operation of taking the adjoint of a linear map between two Hilbert spaces, and since knowing the adjoints of all maps $\mathbb{C} \to H$ is equivalent to knowing the inner product on $H$, it also serves as an axiomatization of the inner product.

\begin{defn} A \emph{\dag\-functor} on a category \cat{C} is a contravariant endofunctor \mbox{$\dag: \cat{C} \to \cat{C}$}, which is the identity on objects and which satisfies $\dag \circ  \dag = \id _\cat{C}$.
\end{defn}

\begin{defn}
\label{dagcategory}
A \emph{\dag\-category} is a category equipped with a particular choice of \mbox{\dag\-functor}.
\end{defn}

\jnoindent
These \dag\-categories have a long history, sometimes going by the name \textit{$*$\-categories}. In particular, they have been well-used in representation theory, especially by Roberts and collaborators \cite{dr89-ndt,lr97-atod} under the framework of C*\-categories, and by others in the study of invariants of topological manifolds~\cite{t94-qik}. They have also been used to study the properties of generalizations of quantum mechanics~\cite{ce08-tqc, v08-ccn}, where it is not assumed that the underlying categories are $\mathbb{C}$-linear. A useful physical intuition is that the \dag\-functor models the time-reversal of processes, and considering it as a fundamental structure gives an interesting new perspective on the development of physical theories~\cite{b06-qq}.

Given a \dag\-category, we denote the action of a \dag\-functor on a morphism $f : A \to B$ as $f ^\dag : B \to A$, and by convention we refer to the morphism $f ^\dag$ as the \emph{adjoint} of $f$. We can now make the following straightforward definitions:

\begin{defn}
In a \dag\-category, a morphism $f : A \to B$ is an \emph{isometry} if $f ^\dag \circ f = \id _A$; in other words, if $f^\dag$ is a retraction of $f$.
\end{defn}

\begin{defn}
In a \dag\-category, a morphism $f:A \to B$ is \emph{unitary} if $f ^\dag \circ f = \id _{A}$ and $f \circ f ^\dag = \id _{B}$; in other words, if $f$ is an isomorphism and $f ^{-1} = f ^\dag$.
\end{defn}

\begin{defn}
In a \dag\-category, a morphism $f:A \to A$ is \emph{self-adjoint} if $f = f ^\dag$.
\end{defn}

\begin{defn}
In a \dag\-category, a morphism $f:A \to A$ is \emph{normal} if $f \circ f ^\dag = f ^\dag \circ f$.
\end{defn}

\subsubsection*{Monoidal categories with duals}

We will work in monoidal categories throughout this paper, and we will require that each object in our monoidal categories has a left and a right dual. In the presence of a \dag\-functor there are then some compatibility equations which we can impose, which we will describe in this section.

There is an important graphical notation for the objects and morphisms in these categories \cite{js91-gtc} which we will rely on heavily. We have already made use of it in the introduction. Objects in a monoidal category are drawn as wires, and the tensor product of two objects is drawn as those objects side-by-side; for consistency with the equation $A \otimes I \simeq A \simeq I \otimes A$, we therefore `represent' the tensor unit object $I$\ as a blank space. Morphisms are represented by `junction-boxes' with input wires coming in underneath and output wires coming out at the top, and composition of morphisms is represented by the joining-up of input and output wires. For visual consistency, the identity morphism on an object is also not drawn. These principles are demonstrated by the following pictures:
\[
\psfrag{a}{\raisebox{-3pt}{$\!A$}}
\psfrag{b}{\raisebox{-3pt}{$\!B$}}
\psfrag{c}{\raisebox{-3pt}{$\!C$}}
\psfrag{f}{$\!g$}
\psfrag{h}{\raisebox{-1pt}{$\!f$}}
\psfrag{g}{\raisebox{-1pt}{$h$}}
{
\setlength\arraycolsep{15pt}
\begin{array}{cccc}
\vc{\includegraphics{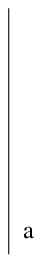}}
&
\vc{\includegraphics{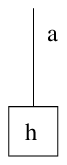}}
&
\vc{\includegraphics{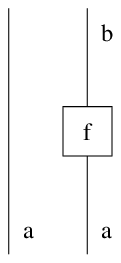}}
&
\vc{\includegraphics{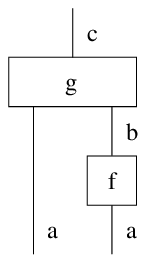}}
\\
\textrm{Object $A$ or}
&
\textrm{Morphism}
&
\textrm{Morphism}
&
\textrm{Morphism}
\\
\textrm{morphism $\id_A$}
&
f: I \to A
&
\id _A \otimes g
&
h \circ (\id _A \otimes g)
\end{array}
}
\]
We will often omit the labels on the wires when it is obvious from the context which object they represent.

We now give the definition of duals, and describe their graphical representation.

\begin{defn}
An object $A$ in a monoidal category has a \emph{left dual} if there exists an object $A^{*_\pL}$ and \emph{left-duality morphisms} $\epsilon ^\pL_A : I \to A^{*_\pL} \otimes A$ and $\eta  ^\pL _A: A \otimes A^{*_\pL} \to I$ satisfying the triangle equations:
\begin{align}
\begin{diagram}[midshaft,nohug,height=25pt,width=35pt,labelstyle=\scriptstyle]
A & &
\\
\dTo <{\id _{A} \otimes \epsilon ^\pL _A} & \rdTo ^ {\id _{A}} &
\\
A \otimes A^* \otimes A & \rTo _{\eta ^\pL _A \otimes \id ^{} _{A}} & A
\end{diagram}
&&
\begin{diagram}[nohug,height=25pt,width=35pt,labelstyle=\scriptstyle]
A^* & &
\\
\dTo <{\epsilon ^\pL _A \otimes \id _{A ^*}} & \rdTo ^{\id _{A^*}} &
\\
A^* \otimes A \otimes A^* & \rTo _{\id _{A ^*} \otimes \eta ^\pL _A} & A^*
\end{diagram}
\end{align}
Analogously, an object $A$ has a \emph{right dual} if there exists an object $A^{*_\pR}$ and \emph{right-duality morphisms} $\epsilon ^\pR _A : I \to A \otimes A^{*_\pR}$ and $\eta ^\pR _A: A^{*_\pR} \otimes A \to I$ satisfying similar equations to those given above.
\end{defn}

\jnoindent
It follows that any two left (or right) duals for an object are canonically isomorphic. To distinguish between the objects $A$ and $A ^{* _\pL}$, we add arrows to our wires, usually drawing an object $A$ with an upward-pointing arrow and drawing $A ^{* _\pL}$ with a downward-pointing one. We use the same notation for $A ^{* _\pR}$, which will not lead to confusion since we will soon choose our duals such that $A ^{* _\pL} = A ^{* _\pR}$ for all objects $A$. We represent the duality morphisms by a `cup' and a `cap' in the following way:
\[
{
\setlength\arraycolsep{20pt}
\begin{array}{cc}
\psfrag{a}{$A$}
\psfrag{d}{\hspace{-14pt}$A ^{* _\pL}$}
\vc{\includegraphics{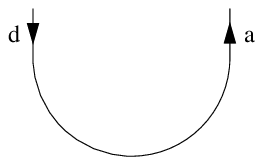}}
&
\psfrag{a}{\hspace{-7pt}$A$}
\psfrag{d}{$A ^{* _\pL}$}
\vc{\includegraphics{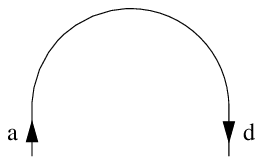}}
\\
\epsilon _A ^\pL : I \to A ^{* _\pL} \otimes A
&
\eta _A ^\pL : A \otimes A ^{* _\pL} \to I
\end{array}
}
\]
The reason for this is made clear by the representation it leads to for the duality equations:
\begin{align*}
\psfrag{x}{\hspace{-5pt}$A$}
\psfrag{y}{\hspace{+0pt}$A$}
\psfrag{d}{$A ^{* _\pL}$}
\vc{\includegraphics{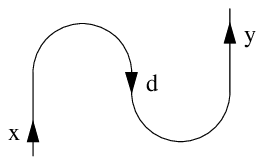}}
=
\psfrag{a}{$A$}
\vc{\includegraphics{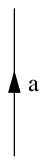}}
&\hspace{20pt}&
\psfrag{x}{\hspace{-14pt}$A ^{* _\pL}$}
\psfrag{a}{$A$}
\psfrag{y}{$A ^{* _\pL}$}
\vc{\includegraphics{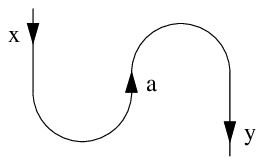}}
=
\psfrag{a}{$A ^{* _\pL}$}
\vc{\includegraphics{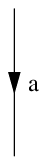}}
\end{align*}
We can therefore `pull kinks straight' in the wires whenever we find them. This is one reason that the graphical representation is so powerful: the eye can easily spot these simplifications, which would be much harder to find in an algebraic representation.
\begin{defn}
A monoidal category \emph{has left duals} (or \emph{has right duals}) if every object $A$ has an assigned left dual $A ^{* _\pL}$ (or a right dual $A ^{* _\pR}$), along with assigned duality morphisms, such that $I ^{* _\pL} = I$ and $(A \otimes B) ^{* _\pL} = B ^{* _\pL} \otimes A ^{* _\pL}$ (or the equivalent with L replaced with R.)
\end{defn}

\jnoindent
The order-reversing property of the $(-) ^{* _\pL}$ and $(-) ^{* _\pR}$ operations for the monoidal tensor product is important: it allows us to choose a dual for $A \otimes B$ given duals of $A$ and $B$ independently. In the presence of a braiding isomorphism $A \otimes B \simeq B \otimes A$ we can suppress this distinction, but this will not be available to us in general.

\begin{defn}
\label{defdualityfunctor}
In a monoidal category with left or right duals, with an assigned left dual for each object or a chosen right dual for each object, the \emph{left duality functor} $(-) ^{ * _\pL}$ and \emph{right duality functor} $(-) ^{*_\pR}$ are contravariant endofunctors that take objects to their assigned duals, and act on morphisms \mbox{$f: A \to B$} in the following way:
\begin{align}
f^{* _\pL} &:= (\id _{A^*} \otimes \eta ^\pL _B) \circ (\id _{A^*} \otimes f \otimes \id _{B ^*}) \circ (\epsilon ^\pL _A \otimes \id _{B^*})
\\
f^{* _\pR} &:= (\eta ^\pR _{B ^*} \otimes \id _{A}) \circ (\id _{B^*} \otimes f \otimes \id _{A ^*}) \circ (\id _{A^*} \otimes \epsilon ^\pR _{A ^*})
\end{align}
\end{defn}

\jnoindent
These definitions can be understood more easily by their pictorial representation:
\begin{equation}
\label{picduals}
\psfrag{fr}{\hspace{-5pt}\raisebox{-1pt}{$f ^{* _\pL}$}}
\psfrag{f}{\raisebox{-1pt}{$\!f$}}
\vc{\includegraphics{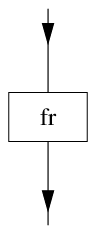}}
:=
\vc{\includegraphics{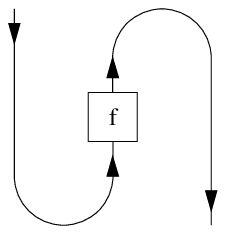}}
\hspace{50pt}
\psfrag{fr}{\hspace{-5pt}\raisebox{-1pt}{$f ^{* _\pR}$}}
\vc{\includegraphics{graphics/dual-right-2}}
:=
\vc{\includegraphics{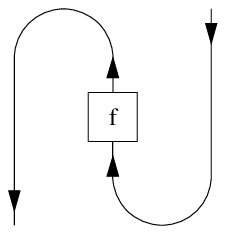}}
\end{equation}

\subsubsection*{Monoidal \dag\-categories with duals}

We now investigate appropriate compatibility conditions in the case that our monoidal category has both duals and a \dag\-functor.

\begin{defn}
A \emph{monoidal \dag\-category} is a monoidal category equipped with a \mbox{\dag\-functor}, such that the associativity and unit natural isomorphisms are unitary. If the monoidal category is equipped with natural braiding isomorphisms, then these must also be unitary.
\end{defn}

\jnoindent
We will not assume that our monoidal categories are strict. A good reference for the essentials of monoidal category theory is \cite{ml97-cwm}.

In a monoidal \dag\-category we can give abstract definitions of some important terminology normally associated with Hilbert spaces.

\begin{defn}
In a monoidal category, the \emph{scalars} are the monoid $\Hom(I,I)$. In a monoidal \dag\-category, the scalars form a monoid with involution.
\end{defn}

\begin{defn}
In a monoidal \dag\-category, a \emph{state} of an object $A$ is a morphism \mbox{$\phi: I \to A$}.
\end{defn}

\begin{defn}
\label{defnorm}
In a monoidal \dag\-category, the \emph{squared norm} of a state $\phi:I \to A$ is the scalar \mbox{$\phi^\dag \circ \phi : I \to I$}.
\end{defn}

\jnoindent
If our \dag\-category also has a zero object, we note that it is quite possible for the squared norm of a non-zero state to be zero. For this reason, as it stands, Definition~\ref{defnorm} seems a poor abstraction of the notion of the squared norm on a vector space. In \cite{v08-ccn} we describe a way to overcome this problem, but it will not affect us here.

Monoidal \dag\-categories have a simpler duality structure than many monoidal categories, as the following lemma shows.

\begin{lemma}
\label{lisr}
In a monoidal \dag\-category, left-dual objects are also right-dual objects. 
\end{lemma}
\jbeginproof
\begin{sloppypar}
Given an object $A$ with a left dual $A ^{* _\pL}$ witnessed by left-duality morphisms \mbox{$\epsilon ^\pL_A : I \to A^{*_\pL} \otimes A$} and \mbox{$\eta  ^\pL _A: A \otimes A^{*_\pL} \to I$}, we can define $\epsilon _A ^\pR := \eta _A ^\pL {}^\dag$ and $\eta _A ^\pL := \epsilon _A ^\pL {}^\dag$ which witness that $A ^{* _\pL}$ is a right dual for $A$.
\qedhere
\end{sloppypar}
\end{proof}

\jnoindent Since left or right duals are always unique up to isomorphism, left duals must be isomorphic to right duals in a monoidal \dag\-category. We will exploit this isomorphism to write $A ^*$ instead of $A ^{* _\pL}$ or $A ^{* _\pR}$, and it follows that $A ^{**} \simeq A$. However, this is not enough to imply that the functors $(-) ^{* _\pL}$ and $(-) ^{* _\pR}$ given in  Definition~\ref{defdualityfunctor} are naturally isomorphic; for this we will require extra compatibility conditions.

\begin{defn}
\label{mondagcatduals}
A \emph{monoidal \dag\-category with duals} is a monoidal \dag\-category such that each object $A$ has an assigned dual object $A^*$ (either left or right by Lemma~\ref{lisr}) with this assignment satisfying \mbox{$(A ^{*}) ^* = A$}, and assigned left and right duality morphisms for each object, such that these assignments are compatible with the \dag\-functor in the following way:
\begin{align}
\epsilon _A ^\pL
= \eta _A ^\pR {}^\dag
= \eta _{A ^*} ^\pL \!{}^\dag
= \epsilon _{A ^*} ^\pR
&&
\eta _A ^\pL
= \epsilon _A ^\pR {}^\dag
= \epsilon _{A ^*} ^\pL \!{}^\dag
= \eta _{A ^*} ^\pR
&&
((-) ^{* _\pL}) ^\dag = ((-) ^\dag) ^{* _\pL}
\end{align}
\end{defn}

\jnoindent
Since the left and right duality morphisms can be obtained from each other using the \mbox{\dag\-functor}, from now on we will only refer directly to the left-duality morphisms, defining $\epsilon ^{}_A := \epsilon _A ^\pL$ and $\eta ^{}_A := \eta _A ^\pL$.

We note that there does not yet exist a precise theorem governing the soundness of the graphical calculus for this precise type of monoidal category with duals, although we fully expect that one could be proved. The graphical calculus used in this paper should therefore be thought of as a shorthand for the underlying morphisms in the category, rather than a calculational method in its own right.

The compatibility condition $((-) ^{* _\pL}) ^\dag = ((-) ^\dag) ^{* _\pL}$ looks asymmetrical, as it does not refer to the right-duality functor $(-) ^{* _\pR}$. We show that it is equivalent to two different compatibility conditions.

\begin{lemma}
\label{equivrequirements}
As a part of the definition of a monoidal \dag\-category with duals, the following compatibility conditions would be equivalent:
\begin{align*}
\mathrm{1. }\,\,\,
((-) ^{* _\pL}) ^\dag = ((-) ^\dag) ^{* _\pL}
&&
\mathrm{2. }\,\,\,
((-) ^{* _\pR }) ^\dag = ((-) ^\dag) ^{* _\pR}
&&
\mathrm{3. }\,\,\,
(-) ^{* _\pL} = (-) ^{* _\pR}
\end{align*}
\end{lemma}

\jbeginproof
From the first two sets of equations between the duality morphisms given in Definition~\ref{mondagcatduals}, it follows directly that $((-) ^{* _\pL}) ^\dag = ((-) ^\dag) ^{* _\pR}$. We combine this with condition 2 above to show that \mbox{$((-) ^{* _\pL}) ^\dag = ((-) ^{* _\pR}) ^\dag$}, and since the \dag\-functor is an involution, it then follows that $(-) ^{* _\pL} = (-) ^{* _\pR}$. Since this argument is reversible we have shown that $2 \Leftrightarrow 3$, and an analogous argument demonstrates that $1 \Leftrightarrow 3$.
\end{proof}

\jnoindent
In a monoidal \dag\-category the three given conditions will therefore all hold, and in particular the functors $(-) ^{* _\pL}$ and $(-) ^{* _\pR}$ will coincide. We denote this unique duality functor as $(-) ^*$. We use conditions 1 for Definition~\ref{mondagcatduals} rather than the more symmetrical definition 3, since it follows from a general `philosophy' of \dag\-categories: wherever sensible, require that structures be compatible with the \dag\-functor.

We can use this result to demonstrate a useful property of the duality functor $(-) ^*$.
\begin{lemma}
In a monoidal \dag\-category with duals, the duality functor $(-) ^*$ is an involution.
\end{lemma}
\jbeginproof
The involution equation is $((-)^* )^* = \id$, and we rewrite this using Lemma~\ref{equivrequirements} as $((-) ^{* _\pL}) ^{* _\pR} = \id$. Writing this out in full, it is easy to demonstrate using the duality equations and the compatibility equations of Definition~\ref{mondagcatduals}.
\end{proof}

\jnoindent
Since the \dag\-functor is also strictly involutive and commutes with the duality functor, their composite is also an involutive functor.
\jnoindent
\begin{defn}
In a monoidal \dag\-category with duals, the \emph{conjugation functor} $(-) _*$ is defined on all morphisms $f$ by $f _* = (f ^*) ^\dag = (f ^\dag) ^*$.
\end{defn}

\jnoindent
Since the \dag\-functor is the identity on objects, we have $A_* = A^*$ for all objects $A$. To make this equality clear we will write $A^*$ exclusively, and the $A_*$ form will not be used.

For any morphism $f:A \to B$ we can use these functors to construct $f_* : A^* \to B^*$, $f ^* : B^* \to A^*$ and $f ^\dag: B \to A$, and it will be important to be able to easily distinguish between these graphically. We will use an approach originally due to Selinger \cite{s07-dccc}, in the form adopted by Coecke and Pavlovic \cite{cp06-qmws}. Given the graphical representation of the duality functor $(-)^*$ given in (\ref{picduals}), we could `pull the kink straight' on the right-hand side of the equation. This would result in a \emph{rotation} of the junction-box for $f$ by half a turn. To make this rotation visible we draw our junction-boxes as wedges, rather than rectangles, breaking their symmetry. The duality $(-)^*$ is given by composing the conjugation functor $(-) _*$ and the \dag\-functor, and since geometrically a half-turn can be built from two successive reflections, this gives us a complete geometrical scheme for describing the actions of our functors:
\[
\psfrag{a}{\hspace{0pt}\raisebox{-0pt}{$A$}}
\psfrag{b}{\hspace{0pt}\raisebox{-0pt}{$B$}}
\psfrag{ad}{\hspace{0pt}\raisebox{-0pt}{$A^*$}}
\psfrag{bd}{\hspace{0pt}\raisebox{-0pt}{$B^*$}}
\psfrag{x}{\hspace{13pt}\raisebox{-0pt}{$A^*$}}
\psfrag{y}{\hspace{13pt}\raisebox{-0pt}{$B^*$}}
\psfrag{f}{\raisebox{-1pt}{$f$}}
\psfrag{fs}{\raisebox{-1pt}{$\!f ^*$}}
\psfrag{fc}{\raisebox{-1pt}{$\!f _*$}}
\psfrag{fd}{\raisebox{-1pt}{$\!f ^\dag$}}
{
\setlength\arraycolsep{10pt}
\begin{array}{cccc}
\vc{\includegraphics{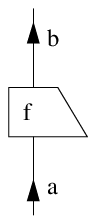}}
&
\vc{\includegraphics{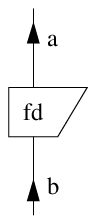}}
&
\vc{\includegraphics{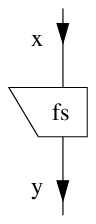}}
&
\vc{\includegraphics{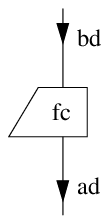}}
\end{array}
}
\]

Our monoidal \dag\-categories with duals are very similar to other structures considered in the literature, such as C*-categories with conjugates~\cite{dr89-ndt, z05-2cc} and strongly-compact-closed categories \cite{ac04-csqp, ac05-apt}. In these contexts the functors $(-)_*$ and $(-)^*$ also play an important role.

\subsubsection*{Involution monoids}

An important tool in functional analysis is the \emph{$*$-algebra}: a complex, associative, unital algebra equipped with an antilinear involutive homomorphism from the algebra to itself which reverses the order of multiplication. Category-theoretically, such a homomorphism is not very convenient to work with, since morphisms in a category of vector spaces are usually chosen to be the {\em linear} maps. However, if the vector space has an inner product, this induces a canonical antilinear isomorphism from the vector space to its dual. Composing this with the antilinear self-involution, we obtain a \emph{linear} isomorphism from the vector space to its dual. This style of isomorphism is much more useful from a categorical perspective, and we use it to define the concept of an \emph{involution monoid}. We will demonstrate that this is equivalent to a conventional \mbox{$*$-algebra} when applied in a category of complex Hilbert spaces. The natural setting for the study of these categorical objects is a category with a conjugation functor, as defined above.

\begin{defn}
In a monoidal category, a \emph{monoid} is an ordered triple $(A,m,u)$ consisting of an object $A$, a \emph{multiplication} morphism \mbox{$m:A \otimes A \to A$} and a \emph{unit} morphism \mbox{$u : I \to A$}, which satisfy \emph{associativity} and \emph{unit} equations:
\begin{equation}
\vc{\includegraphics{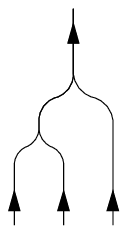}}
=
\vc{\includegraphics{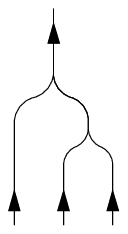}}
\hspace{30pt}
\vc{\includegraphics{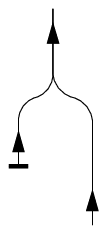}}
=
\vc{\includegraphics{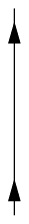}}
=
\vc{\includegraphics{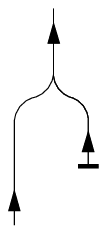}}
\end{equation}
\end{defn}

\begin{defn}
In a monoidal \dag\-category with duals, an \emph{involution monoid} $(A,m,u;s)$ is a monoid $(A,m,u)$ equipped with a morphism $s:A \to A^*$ called the \emph{linear involution}, which is a morphism of monoids with respect to the monoid structure $(A^*, m_*, u_*)$ on $A^*$, and which satisfies the \emph{involution condition}
\begin{equation}
s_* \circ s = \id _{A}.
\end{equation}
\end{defn}

\jnoindent It follows from this definition that $s$ and $s _*$ are mutually inverse morphisms, since applying the conjugation functor to the involution condition gives $s \circ s_* = \id _{A ^*}$. We also note that for any such involution monoid $s: A \to A^*$ and $s^*:A \to A^*$ are parallel morphisms, but they are not necessarily the same. 

\begin{defn}
\label{morphinvmon}
\begin{sloppypar}
In a monoidal \dag\-category with duals, given involution monoids $(A,m,u; s_A)$ and $(B,n,v; s_B)$, a morphism $f:A \to B$ is a \emph{homomorphism of involution monoids} if it is a morphism of monoids, and if it satisfies the \emph{involution-preservation condition}
\begin{equation}
s_B \circ f = f_* \circ s_A.
\end{equation}
\end{sloppypar}
\end{defn}

\jnoindent
If an object $B$ is self-dual, it is possible for the involution $s_B : B \to B$ to be the identity. Let $(B,n,v; \id _{B})$ be such an involution monoid. In this case, it is sometimes possible to find an embedding $f : (A,m,u; s_A) \into (B, n, v; \id _{B})$ of involution monoids even when the linear involution $s_A$ is \emph{not} trivial! We will see an example of this in the next section.

The following lemma establishes that the traditional concept of $*$-algebra and the categorical concept of an involution monoid are the same, in an appropriate context. We demonstrate the equivalence for finite-dimensional algebras, since the category of finite-dimensional complex vector spaces forms a category with duals. However, involution monoids are useful far more generally, and with a careful choice of conjugation functor could be used just as well to describe infinite-dimensional algebras with an involution.

\begin{lemma}
\label{invollemma}
For a unital, associative algebra on a finite-dimensional complex Hilbert space $V$, there is a correspondence between the following structures:
\jbeginenumerate
\item antilinear maps $t: V \to V$  which are involutions, and which are order-reversing algebra homomorphisms;
\item linear maps $s: V \to V^*$ where $V^*$ is the dual space of $V$, satisfying $s_* \circ s = \id _{V}$, and which are algebra homomorphisms to the conjugate algebra on $V ^*$.
\end{enumerate}
\vspace{-5pt}
Furthermore, the natural notions of homomorphism for these structures are also equivalent.
\end{lemma}

\jbeginproof
We first deal with the implication $1 \Rightarrow 2$. We construct the linear isomorphism $s$ by defining $s \circ \phi := \jbigl t(\phi) \jbigr _*$ for an arbitrary morphism $\phi: \mathbb{C} \to V$. This is linear, because both $t$ and $(-) _*$ are antilinear. It is a map \mbox{$V \to V^*$} since $t(\phi)$ is an element of $V$, and the complex conjugation functor $(-) _*$ takes $V$ to $V^*$. Checking the identity $s_* \circ s = \id _{V}$, we have
\[
s_* \circ s \circ \phi = s _* \circ \jbigl t(\phi) \jbigr _* = \jbigl s \circ t(\phi) \jbigr _* = \jbigl tt(\phi) \jbigr _{**} = \phi.
\]
The monoid homomorphism condition is demonstrated similarly, for arbitrary states $\phi$ and $\psi$ of $V$: 
\begin{align*}
s \circ m \circ (\phi \otimes \psi) &= \jbigl t(m \circ (\phi \otimes \psi)) \jbigr _*
\proofnote{definition of $s$}
\\
&= \jbigl m \circ (t\psi \otimes t\phi) \jbigr _*
\proofnote{$t$ is order-reversing homomorphism}
\\
&= m _* \circ \jbigl (t\phi) _* \otimes (t \psi) _* \jbigr _{\phantom{*}}
\proofnote{order-reversing functoriality of $(-) _*$}
\\
&= m _* \circ \jbigl s\phi \otimes s\psi \jbigr
\proofnote{definition of $s$}
\\[5pt]
s \circ u &= \jbigl t(u) \jbigr _* = u _*
\proofnote{definition of $s$, $t$ is homomorphism}
\end{align*}
For the implication $2 \Rightarrow 1$, we define $t(\phi) := (s \circ \phi) _*$ for all elements $\phi$ of $V$. The proof that $t$ has the required properties is similar to the proof involved in the implication $1 \Rightarrow 2$. The constructions of $s$ and $t$ in terms of each other are clearly inverse, and so the equivalence has been demonstrated.

We now check that homomorphisms between these structures are the same. Our notion of homomorphism between structures of type 2 is given by that in Definition~\ref{morphinvmon}, and there is a natural notion of homomorphism between monoids equipped with an antilinear self-involution. Consider algebras $(A,m,u)$ and $(B,n,v)$ equipped with antilinear involutive order-reversing homomorphisms \mbox{$t_A: A \to A$} and $t_B : B {\to} B$ respectively, and let \mbox{$f:A \to B$} be any continuous linear map. It will be compatible with the involutions if $t_B \circ f = f \circ t_A$. Acting on some state $\phi$ of $A$, and constructing linear maps $s_A : A \to A^*$ and $s_B : B \to B^*$ in the manner defined above, we obtain $t_B \circ f \circ \phi = {s_B} _* \circ (f \circ \phi) _* = {s_B} _* \circ f_* \circ \phi_*$ and $f \circ t_A \circ \phi = f \circ {s_A} _* \circ \phi _*$. Equating these and complex-conjugating we have $s_B \circ f = f_* \circ s_A$ as required. Conversely, let $(A,m,u; s_A)$ and $(B,n,v; s_B)$ be involution monoids in \cat{Hilb}, and let $f:A \to B$ again be any linear map. If the involution-preservation condition $s_B \circ f = f_* \circ s_A$ holds, then applying an arbitrary state $\phi$ we obtain $s_B \circ f \circ \phi = \jbigl t(f \circ \phi) \jbigr {}_*$ and $f_* \circ s_A \circ \phi = f_* \circ (t\phi) _*$ respectively for the left and right sides of the equation. Equating these and complex-conjugating, we obtain $t(f \circ \phi) = f \circ (t \phi)$ as required.
\end{proof}

\section{Results on \dag\-Frobenius monoids}
\label{frobmonsec}

\subsubsection*{Introducing \dag\-Frobenius monoids}

We begin with definitions of the important concepts.

\begin{defn}
In a monoidal category, a \emph{comonoid} is the dual concept to a monoid; that is, it is an ordered triple $(A,n,v) _\times$ consisting of an object $A$, a \emph{comultiplication} \mbox{$n:A \to A \otimes A$} and a \emph{counit} $v:A \to I$, which satisfy \emph{coassociativity} and \emph{counit} equations:
\begin{equation}
\vc{\includegraphics{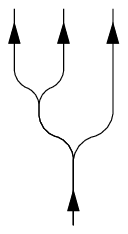}}
=
\vc{\includegraphics{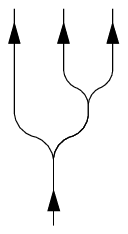}}
\hspace{30pt}
\vc{\includegraphics{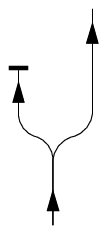}}
=
\vc{\includegraphics{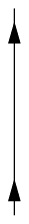}}
=
\vc{\includegraphics{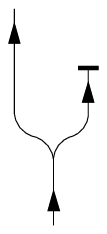}}
\end{equation}
\end{defn}

\jnoindent
If an object has both a chosen monoid structure and a chosen comonoid structure, then there is an important way in which these might be compatible with each other.

\begin{defn}
In a monoidal category, a \emph{Frobenius structure} is a choice of monoid $(A,m,u)$ and comonoid $(A,n,v) _\times$ for some object $A$, such that the multiplication $m$ and the comultiplication $n$ satisfy the following equations:
\begin{equation}
\vc{\includegraphics{graphics/small-frobenius2}}
=
\vc{\includegraphics{graphics/small-frobenius3}}
=
\vc{\includegraphics{graphics/small-frobenius1}}
\end{equation}
\end{defn}

\jnoindent
Reading these diagrams from bottom to top, the splitting of a line represents the comultiplication $n$, and merging of two lines represents the multiplication $m$.

This geometrical definition of a Frobenius structure, although well-known, is superficially quite different to the `classical' definition in terms of an exact pairing. The equivalence of these two definitions was first observed by Abrams \cite{a96-2dtqft}, and an accessible discussion of the different possible ways to define a Frobenius algebra is given in the book by Kock \cite{k04-fa2d}. This geometrical definition was first suggested by Lawvere,  and was subsequently popularized in the lecture notes of Quinn~\cite{q95-l}. An important property of a Frobenius structure is that it can be used to demonstrate that the underlying object is self-dual.

If we are working in a \dag\-category, from any monoid $(A,m,u)$ we can canonically obtain an `adjoint' comonoid $(A,m^\dag, u ^\dag) _\times$, and it is then natural to make the following definition.

\begin{defn}
\label{defdagfrob}
In a monoidal \dag\-category, a monoid $(A,m,u)$ is a \emph{\dag\-Frobenius monoid} if it forms a Frobenius structure with its adjoint $(A,m^\dag, u ^\dag) _\times$.
\end{defn}
\jnoindent
This construction is similar to an \emph{abstract Q\-systems}~\cite{lr97-atod}. Given a \dag\-Frobenius monoid $(A,m,u)$, we refer to $m^\dag$ as its comultiplication and to $u ^\dag$ as its counit. 

\subsubsection*{Involutions on \dag\-Frobenius monoids}

We now look at the relationship between \mbox{\dag\-Frobenius} monoids and the involution monoids of Section~\ref{strucdagcat}. We will see that a \dag\-Frobenius monoid can be given the structure of an involution monoid in two canonical ways, which in general will be different. 

\begin{defn}
\label{definv}
In a monoidal \dag\-category with duals, a \dag\-Frobenius monoid $(A,m,u)$ has a \emph{left involution} $s_\pL: A \to A^*$ and \emph{right involution} $s_\pR : A \to A^*$ defined as follows:
\begin{equation}
\label{canonicalinv}
\begin{array}{ccc}
{
\hspace{-20pt}
\vc{\includegraphics{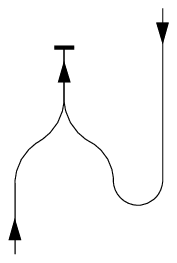}}
\hspace{-20pt}
}
=
{
\hspace{-20pt}
\vc{\includegraphics{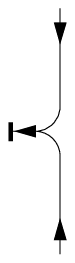}}
\hspace{-20pt}
}
&&
{
\hspace{-20pt}
\vc{\includegraphics{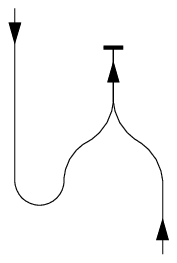}}
\hspace{-20pt}
}
=
{
\hspace{-20pt}
\vc{\includegraphics{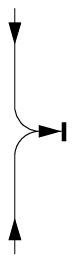}}
\hspace{-20pt}
}
\\
s_\pL := \jbigl (u ^\dag \circ m) \otimes \id _{A ^*} \jbigr \circ \jbigl \id _{A} \otimes \epsilon _{A ^*} \jbigr
&&
s_\pR := \big( \id _{A ^*} \otimes (u ^\dag \circ m) \big) \circ \big( \epsilon _A \otimes \id _{A} \big)
\end{array}
\end{equation}
\end{defn}

\jnoindent
In each case the second picture is just a convenient shorthand, which should literally be interpreted as the first picture. These involutions interact with the conjugation and transposition functors in interesting ways, as we explore in the next lemma.

\begin{lemma}
\label{invprop}
In a monoidal \dag\-category with duals, the left and right involutions of a \mbox{\dag\-Frobenius} monoid satisfy the following equations:
\begin{align}
\label{lri1}
s_\pL {}^* = s_\pR \hspace{1pt}, \,
\hspace{3.6pt}
\,&\, s_\pR {}^* = s_\pL
\\
\label{lri2}
s_\pL {}_* = s _\pL ^{-1},
\,\,&\,
s_\pR {}_* = s _\pR ^{-1}
\\
\label{lri3}
s_\pL ^{-1} = s_\pR {}^\dag, \,
\hspace{0.4pt}
\,&\, s_\pR ^{-1} = s _\pL {} ^\dag
\end{align}
\end{lemma}

\jbeginproof
The equations (\ref{lri1}) follow from the definitions of the involutions and the graphical representation of the functor $(-) ^*$, which rotates a diagram half a turn about an axis perpendicular to the page. The equations (\ref{lri2}) follow from the \dag\-Frobenius and unit equations; taking the right-involution case, we show this by establishing that \mbox{$s_\pR {}_* \circ s_\pR = \id _{A}$} with the following graphical proof:
\[
{
\hspace{-10pt}
\vc{\includegraphics{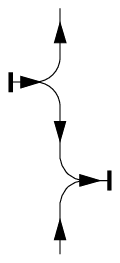}}
\hspace{-10pt}
}
=
{
\hspace{-10pt}
\vc{\includegraphics{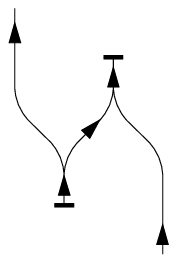}}
\hspace{-10pt}
}
=
{
\hspace{-10pt}
\vc{\includegraphics{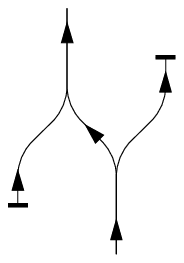}}
\hspace{-10pt}
}
=
{
\hspace{-10pt}
\vc{\includegraphics{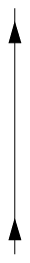}}
\hspace{-10pt}
}
\]
Applying the functor $(-)_*$ to this equation gives $s_\pR \circ s_\pR {}_* = \id _{A^*}$, establishing that $s_\pR$ and $s_\pR {}_*$ are inverse; applying the functor $(-) ^*$ to this argument establishes that $s_\pL$ and $s_\pL ^*$ are inverse. The equations (\ref{lri3}) follow from the equations (\ref{lri1}) and (\ref{lri2}) and the properties of the functors $(-)^*$, $(-) _*$ and $\dag$.
\end{proof}

\jnoindent
We note that left and right involutions could be defined for arbitrary monoids in a monoidal \dag\-category with duals, but they would not satisfy equations (\ref{lri2}) and (\ref{lri3}) above.

We now combine these results on involutions of \dag\-Frobenius monoids with the concept of an involution monoid from Section~\ref{strucdagcat}.

\begin{lemma}
\label{dagfrobinvol}
In a monoidal \dag\-category with duals, given a \dag\-Frobenius monoid $(A,m,u)$ we can canonically obtain two involution monoids $(A,m,u;s_\pL)$ and $(A,m,u; s_\pR)$, where $s_\pL$ and $s_\pR$ are respectively the left and right involutions associated to the monoid.
\end{lemma}
\jbeginproof
We deal with the right-involution case; the left-involution case is analogous. We must show that $s_R: A \to A_*$ is a morphism of monoids, and that it satisfies the involution condition. We first show that it preserves multiplication, employing the Frobenius, unit and associativity laws:
\[
{
\hspace{-20pt}
\vc{\includegraphics{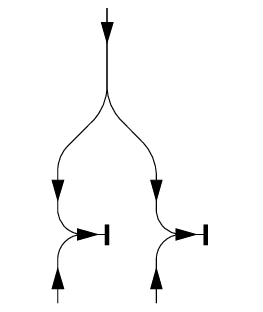}}
\hspace{-20pt}
}
=
{
\hspace{-20pt}
\vc{\includegraphics{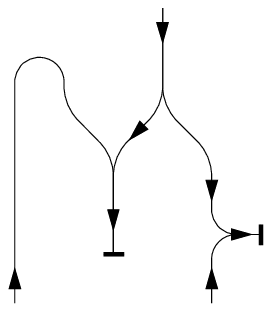}}
\hspace{-20pt}
}
=
{
\hspace{-20pt}
\vc{\includegraphics{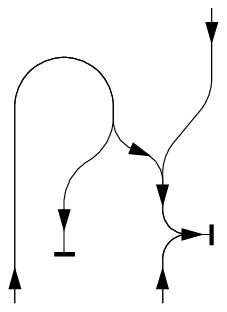}}
\hspace{-20pt}
}
=
{
\hspace{-20pt}
\vc{\includegraphics{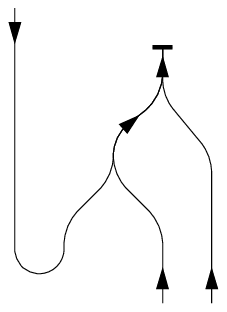}}
\hspace{-20pt}
}
=
{
\hspace{-20pt}
\vc{\includegraphics{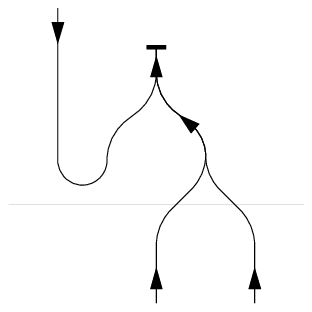}}
\hspace{-20pt}
}
=
{
\hspace{-20pt}
\vc{\includegraphics{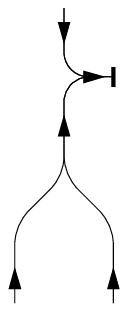}}
\hspace{-20pt}
}
\]
We omit the proof that $s_R$ preserves the unit, as it is straightforward. The involution condition  $s_R {}_* \circ s_R = \id _{A}$ follows from one of the equations (\ref{lri2}) in Lemma~\ref{invprop}.
\end{proof}

\jnoindent
This leads us to the following definition.

\begin{defn}
In a monoidal \dag\-category with duals, a \emph{\dag\-Frobenius left- \emph{(or \emph{right-})} involution monoid} is an involution monoid $(A,m,u; s)$ such that the monoid $(A,m,u)$ is \dag\-Frobenius, and such that the involution $s$ is the left (or right) involution of the \mbox{\dag\-Frobenius} monoid in the manner described by Definition~\ref{definv}.
\end{defn}
\jnoindent
A homomorphism of \dag\-Frobenius left- or right-involution monoids would therefore be required to preserve the involution as well as the multiplication and unit, as per Definition~\ref{morphinvmon}.

A useful property of \dag\-Frobenius right-involution monoids is described by the following lemma, which gives a necessary and sufficient algebraic condition for a monoid homomorphism to be an isometry.
\begin{lemma}
\label{counitunitary}
In a monoidal \dag\-category with duals, a homomorphism of \mbox{\dag\-Frobenius} right-involution monoids is an isometry if and only if it preserves the counit.
\end{lemma}
\jbeginproof
Let $j:(A,m,u) \to (B,n,v)$ be a homomorphism between \dag\-Frobenius right-involution monoids. Assuming that $j$ preserves the counit, we show that it is an isometry by the following graphical argument. The third step uses the fact that $j$ preserves the involution, the fifth that it is a homomorphism of monoids, and the sixth that it preserves the counit.
\begin{gather*}
\psfrag{j}{\raisebox{-0.5pt}{\,$j$}}
\psfrag{jd}{\raisebox{-1pt}{$j ^\dag$}}
\psfrag{js}{\raisebox{-0.5pt}{\!$j _*$}}
{
\hspace{-10pt}
\vc{
\includegraphics{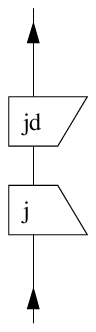}}
}
=
{
\vc{\includegraphics{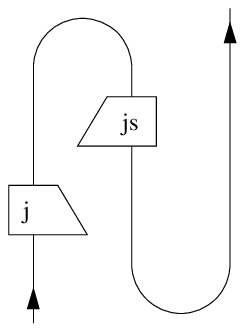}}
}
=
{
\vc{\includegraphics{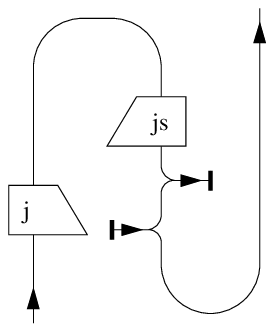}}
}
=
{
\vc{\includegraphics{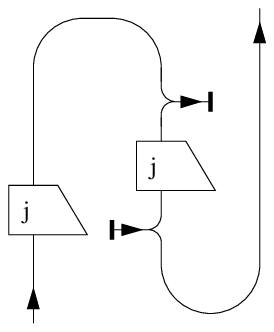}}
}
\\
\psfrag{j}{$j$}
\psfrag{jd}{$j ^\dag$}
\psfrag{js}{$j _*$}
\hspace{100pt}
=
{
\vc{\includegraphics{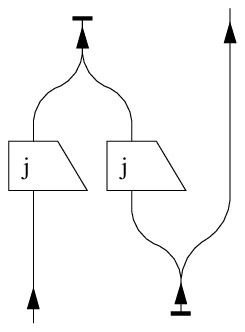}}
}
=
{
\vc{\includegraphics{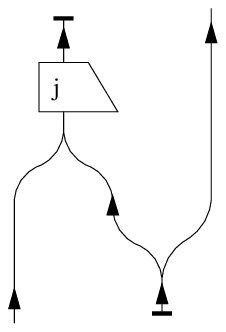}}
}
=
{
\vc{\includegraphics{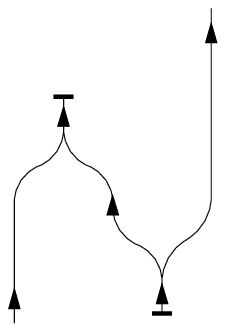}}
}
=
{
\vc{\includegraphics{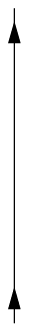}}
}
\end{gather*}
Now instead assume that $j$ is an isometry. It is a homomorphism, so we have the unit-preservation equation $j \circ  u = v$, and therefore $j ^\dag \circ j \circ u = u = j ^\dag \circ v$. Applying the $\dag$-functor to this we obtain $u ^\dag = v ^\dag \circ j$, which is the counit-preservation condition.
\end{proof}

\subsubsection*{Special unitary \dag\-Frobenius monoids}

We will mostly be interested in the case when the two involutions are the same, and we now explore under what conditions this holds.

\begin{defn}
In a monoidal \dag\-category with duals, a \dag\-Frobenius monoid is \emph{unitary} if the left involution, or equivalently the right involution, is unitary.
\end{defn}
\jnoindent
That these two conditions are equivalent follows from Lemma~\ref{invprop}.
\begin{defn}
In a braided monoidal \dag\-category with duals, a \dag\-Frobenius monoid is \emph{balanced-symmetric} if the following equation is satisfied:
\begin{equation}
{
\hspace{-10pt}
\vc{\includegraphics{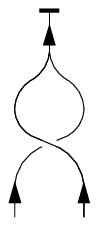}}
\hspace{-10pt}
}
=
{
\hspace{-10pt}
\vc{\includegraphics{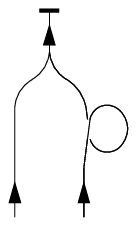}}
\hspace{-10pt}
}
\end{equation}
\end{defn}

\jnoindent
The term \emph{symmetric} is standard (for example, see \cite[Section~2.2.9]{k04-fa2d}), and describes a similar property that lacks the `balancing loop' on one of the legs of the right-hand side of the equation. In \cat{Hilb} this loop is the identity and so the concepts are the same, but this may not be the case in other categories of interest.

\begin{lemma}
\label{unitarylemma}
In a monoidal \dag\-category with duals, the following properties of a \mbox{\dag\-Frobenius} monoid are equivalent:
\jbeginenumerate
\item it is unitary;
\item it is balanced-symmetric;
\item the left and right involutions are the same;
\end{enumerate}
\jnoindent
where property 2 only applies if the monoidal structure has a braiding.
\end{lemma}

\jbeginproof
We first give a graphical proof that $3 \Rightarrow 2$, using property 3 to transform the second expression into the third:
\vspace{-10pt}
\[
{
\hspace{-10pt}
\vc{\includegraphics{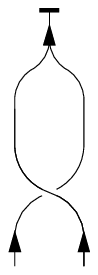}}
\hspace{-10pt}
}
=
{
\hspace{-10pt}
\vc{\includegraphics{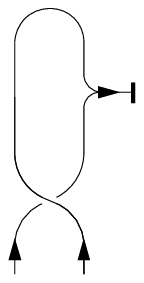}}
\hspace{-10pt}
}
=
{
\hspace{-10pt}
\vc{\includegraphics{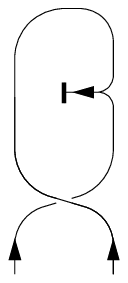}}
\hspace{-10pt}
}
=
{
\hspace{-10pt}
\vc{\includegraphics{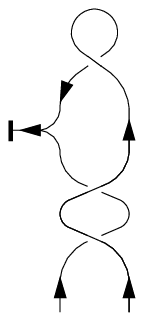}}
\hspace{-10pt}
}
=
{
\hspace{-10pt}
\vc{\includegraphics{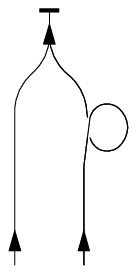}}
\hspace{-10pt}
}
\]
A similar argument shows that $2 \Rightarrow 3$. From equations (\ref{lri3}) of Lemma~\ref{invprop} it follows that $1 \Leftrightarrow 3$, and so all three properties are equivalent.
\end{proof}

\jnoindent
We will mostly use the term `unitary' to refer to these equivalent properties, since it is more obviously in keeping with the general philosophy of \dag\-categories, that all structural isomorphisms should be unitary. We also note that if a \dag\-Frobenius left- or right-involution monoid is unitary then we can simply refer to it as a `\dag\-Frobenius involution monoid', as the left and right involutions coincide in that case.

One particularly nice feature of unitary \dag\-Frobenius monoids is that we can canonically obtain an abstract `dimension' of their underlying space from the multiplication, unit, comultiplication and counit, as the following lemma shows. In a category of vector spaces and linear maps, this dimension will correspond to the dimension of the vector space.
\begin{defn}
In a monoidal \dag\-category with duals, the \emph{dimension} of an object $A$ is given by the scalar $\epsilon_A {}^\dag \circ \epsilon_A : I \to I$, and is denoted $\dim(A)$.
\end{defn}
\begin{lemma}
\label{dimlemma}
In a monoidal \dag\-category with duals, given a unitary \dag\-Frobenius monoid $(A,m,u)$, $\dim(A) = u ^\dag \circ m \circ m ^\dag \circ u$; that is, the dimension of $A$ is equal to the squared norm of $m ^\dag \circ u$. Also, $\dim(A) = \dim(A) ^*$.
\end{lemma}
\jbeginproof
We demonstrate this with the following series of pictures:
\[
\dim(A)
=
\hspace{-10pt}
\vc{\includegraphics[scale=0.9]{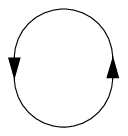}}
\hspace{-10pt}
=
\hspace{-15pt}
\vc{\includegraphics[scale=0.9]{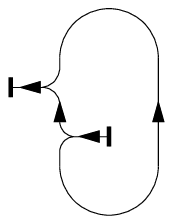}}
\hspace{-10pt}
=
\hspace{-10pt}
\vc{\includegraphics[scale=0.9]{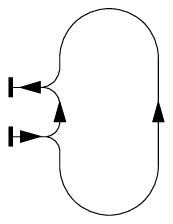}}
\hspace{-10pt}
=
\hspace{-10pt}
\vc{\includegraphics[scale=0.9]{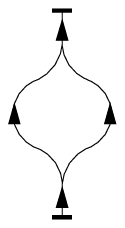}}
\hspace{-10pt}
=
\hspace{-10pt}
\vc{\includegraphics[scale=0.9]{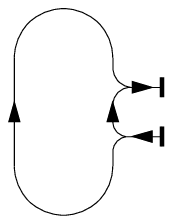}}
\hspace{-10pt}
=
\hspace{-10pt}
\vc{\includegraphics[scale=0.9]{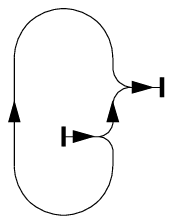}}
\hspace{-15pt}
=
\hspace{-10pt}
\vc{\includegraphics[scale=0.9]{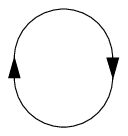}}
\hspace{-10pt}
=
\dim(A) ^*
\]
The central diagram is $u ^\dag \circ m \circ m ^\dag \circ u$, so this proves the lemma.
\end{proof}

\jnoindent
The notion of the dimension of an object is a crucial one in the theory of monoidal categories with duals, and is studied in depth throughout the literature \cite{bw99-sc, dr89-ndt, lr97-atod}. However, we do not rely on it heavily in this paper, and more axioms would be required for our category than those assumed here for the dimension to have good properties, such as being independent of the choice of duality morphisms, or being an element of the integers.

We now introduce one final property of a \dag\-Frobenius monoid.

\begin{defn}
In a monoidal \dag\-category, a \dag\-Frobenius monoid $(A,m,u)$ is \emph{special} if $m  \circ m^\dag = \id _{A}$; that is, if the comultiplication is an isometry.
\end{defn}
\jnoindent
The term \emph{special} goes back to Quinn~\cite{q95-l}. A special \dag\-Frobenius monoid is the same as an \emph{abstract Q-system}~\cite{lr97-atod}, and a useful lemma proved in that reference is that if a monoid $(A,m,u)$ satisfies $m \circ m ^\dag = \id _A$, then it is necessarily a special \dag\-Frobenius monoid.

It simplifies the expression for the dimension of the underlying space, as demonstrated by this lemma.
\begin{lemma}
\label{dimnormu}
In a monoidal \dag\-category with duals, a special unitary \dag\-Frobenius monoid $(A,m,u)$ has $\dim(A) = u ^\dag \circ u$; that is, the dimension of $A$ is equal to the squared norm of $u$.
\end{lemma}
\jbeginproof
Straightforward from Lemma~\ref{dimlemma}.
\end{proof}

\subsubsection*{Endomorphism monoids}

Given any Hilbert space $H$, it is often useful to consider the algebra of bounded linear operators on $H$. These give the prototypical examples of C*\-algebras, with the \mbox{$*$-involution} given by taking the operator adjoint. In a monoidal category with duals we can construct \emph{endomorphism monoids}, which are categorical analogues of these algebras of bounded linear operators. These well-known constructions, which go back at least to M\"uger~\cite{m03-sct1}, form an important class of \dag\-Frobenius monoids, and that they have particularly nice properties.

\begin{defn}
In a monoidal category, for an object $A$ with a left dual $A^{* _\pL}$, the \emph{endomorphism monoid} $\End(A)$ is defined by 
\begin{equation}
\End(A) := \big( A^{* _\pL} \otimes A, {\id _{A^{* _{\pL}}}} \otimes {\eta ^\pL _A} \otimes {\id _{A}}, \epsilon ^\pL _A \big).
\end{equation}
\end{defn}

The following lemma describes a well-known connection between categorical duality and Frobenius structures.
\begin{lemma}
\label{endoisfrob}
In a monoidal \dag\-category with duals, an endomorphism monoid is a \mbox{\dag\-Frobenius} monoid.
\end{lemma}

\jbeginproof
That the \dag\-Frobenius property holds for an endomorphism monoid $\End(A)$ is clear from its graphical representation, which we give here:
\[
{
\hspace{-10pt}
\vc{\includegraphics{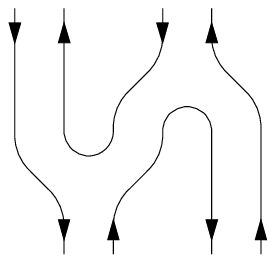}}
\hspace{-10pt}
}
=
{
\hspace{-10pt}
\vc{\includegraphics{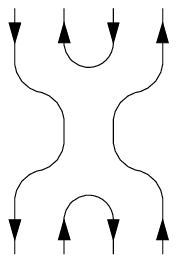}}
\hspace{-10pt}
}
=
{
\hspace{-10pt}
\vc{\includegraphics{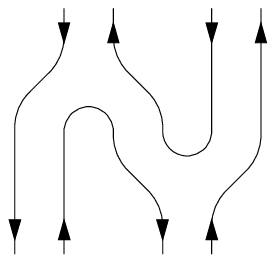}}
\hspace{-10pt}
}
\lowqedhere{50pt}
\]
\end{proof}

\jnoindent
They are examples of the unitary monoids discussed in the previous section.

\begin{lemma}
\label{endoequalinv}
In a monoidal \dag\-category with duals, endomorphism monoids are unitary. 
\end{lemma}

\jbeginproof
Following equation (\ref{canonicalinv}) for the left involution associated to a \dag\-Frobenius monoid, we obtain the following:
\[
\includegraphics{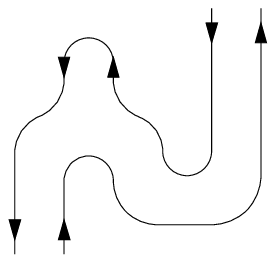}
\]
This is clearly the identity on $A^* \otimes A$. The right involution is also the identity, by the conjugate of this picture. By Lemma~\ref{unitarylemma} the \dag\-Frobenius monoid must therefore unitary.
\end{proof}

\jnoindent
We note that the order-reversing property of the duality functor $(-)^*$ is crucial here, as the only canonical choice of `identity' morphism $A ^* \otimes A \to A \otimes A^*$ would be the braiding isomorphism, but such a braiding is not necessarily present. Also, although the linear involution associated with an endomorphism monoid is the identity, the induced order-reversing \emph{antilinear} involution on $A^* \otimes A$ is certainly not the identity: it is given by taking the name of an operator to the name of the adjoint to that operator, as can be checked by going through the correspondence described in Lemma~\ref{invollemma}.

The following lemma is a formal description of the intuitive notion that an algebra should have a homomorphism into the algebra of operators on the underlying space, given by taking the right action of each element.
\begin{lemma}
\label{embedmonoidlemma}
Let $(A,m,u)$ be a monoid in a monoidal category in which the object $A$ has a left dual. Then $(A,m,u)$ has a monic homomorphism into the endomorphism monoid of $A$.
\end{lemma}

\jbeginproof
The embedding morphism $h : (A,m,u) \into \End(A)$ is defined by
\begin{equation}
h := (\id _{A ^*} \otimes m ) \circ (\epsilon ^\pL _A \otimes \id _{A}),
\end{equation}
which has the following graphical representation:
\begin{equation*}
\begin{diagram}[height=30pt]
A^* \otimes A
\\
\uTo < h
\\
A
\end{diagram}
=
\vc{\includegraphics{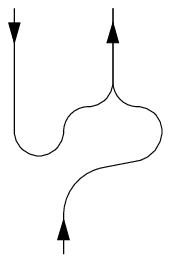}}
\end{equation*}
We show that it is monic by postcomposing with $u ^* \otimes \id _{A}$, which acts as a retraction:
\[
{
\vc{\includegraphics{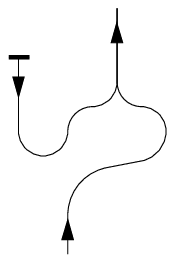}}
}
=
{
\vc{\includegraphics{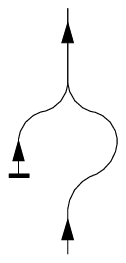}}
}
=
{
\vc{\includegraphics{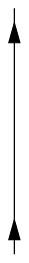}}
}
\]
Next we show that $h$ preserves the multiplication operation, employing a duality equation and the associative law:
\[
{
\vc{\includegraphics{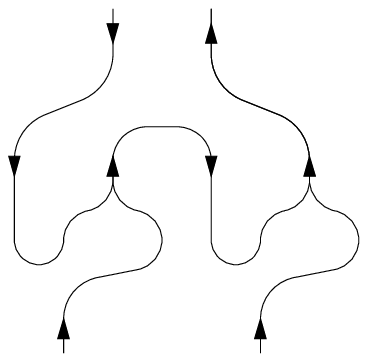}}
}
=
{
\vc{\includegraphics{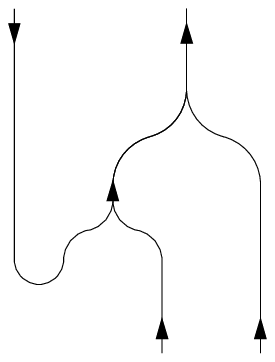}}
}
=
{
\vc{\includegraphics{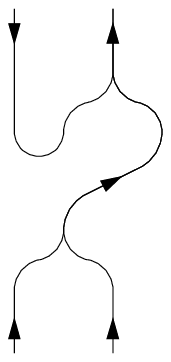}}
}
\]
Finally, we show that the embedding preserves the unit, employing the unit law:
\[
{
\vc{\includegraphics{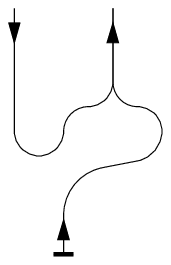}}
}
=
{
\vc{\includegraphics{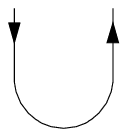}}
}
\eqno{\droptext{48pt}{\qedhere}}
\]
\end{proof}

However, as we saw in the introduction, for the case of \dag\-Frobenius monoids this embedding has a special property: it preserves an involution. We establish this formally in the following lemma.
\begin{lemma}
\label{involembedding}
Let $(A,m,u; s_R)$ be a \dag\-Frobenius right-involution monoid. Then the canonical embedding of $(A,m,u;s_R)$ into the \dag\-Frobenius involution monoid $\End(A)$ is a morphism of involution monoids.
\end{lemma}
\jbeginproof
By Lemma~\ref{embedmonoidlemma} the embedding must be a morphism of monoids. Note that we do not need to specify whether we are using the left or right involution of $\End(A)$, since by Lemma~\ref{endoequalinv} they are both the identity. We must show that this embedding morphism $k : A \into A^* \otimes A$ satisfies the involution condition $k = k_* \circ s_R$ given in Definition~\ref{morphinvmon}. The proof uses the Frobenius law and the unit law.
\[
{
\hspace{-10pt}
\vc{\includegraphics{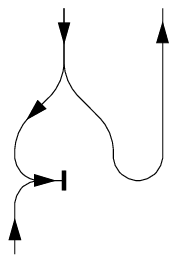}}
\hspace{-10pt}
}
=
{
\hspace{-10pt}
\vc{\includegraphics{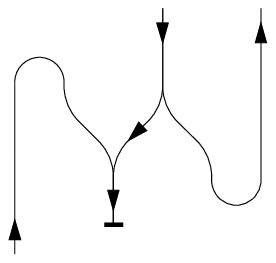}}
\hspace{-10pt}
}
=
{
\hspace{-10pt}
\vc{\includegraphics{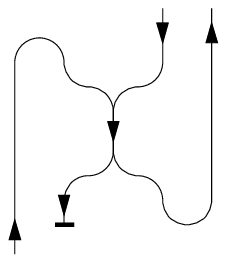}}
\hspace{-10pt}
}
=
{
\hspace{-10pt}
\vc{\includegraphics{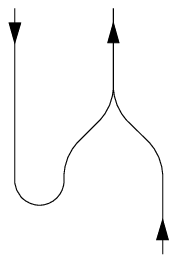}}
\hspace{-10pt}
}
\eqno{\droptext{48pt}{\qedhere}}
\]
\end{proof}

\jnoindent
It is worth noting that a symmetry has been broken; this lemma would not hold with `right-involution' replaced with `left-involution'. This is a consequence of defining the underlying object of our endomorphism monoid to be $A^* \otimes A$ rather than $A \otimes A^*$. In a braided monoidal category there would be no essential difference, but we are working at a higher level of generality.

\subsubsection*{An embedding lemma}

We finish this section by demonstrating another general property of \dag-Frobenius involution algebras. Just as every involution-closed subalgebra of a finite-dimensional C*\-algebra is also a C*-algebra, we will show that every involution-closed submonoid of a \dag\-Frobenius involution monoid is also \dag\-Frobenius. The next section makes this analogy with C*\-algebras precise, but we can prove it here as a general result about \dag\-Frobenius algebras.

\begin{lemma}
\label{subalgisfrob}
In a monoidal \dag\-category with duals, let $(A,m,u; s)$ be an involution monoid with an involution-preserving \dag\-embedding into a \dag\-Frobenius left- (or right-) involution monoid. Then $(A,m,u;s)$ is itself a \mbox{\dag\-Frobenius} left- (or right-) involution monoid. 
\end{lemma}
\jbeginproof
We will deal with the left-involution case; the right-involution case is analogous. Let \mbox{$p : (A,m,u; s) \into (B,n,v; t)$} be a \dag\-embedding of an involution monoid into a \mbox{\dag\-Frobenius} left-involution monoid. The \dag\-embedding property means that $p^\dag \circ p = \id _{A}$. In our graphical representation we will use a thin line for $A$ and a thick line for $B$, and a transition between these types of line for the embedding morphism $p$. The involution-preservation condition $t \circ p = p _* \circ s$ is then represented by the following picture:
\[
{
\hspace{-10pt}
\psfrag{p}{$\,p$}
\vc{\includegraphics{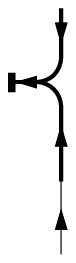}}
\hspace{-10pt}
}
=
{
\hspace{-10pt}
\psfrag{ps}{$\,\,\,p _*$}
\psfrag{s}{\raisebox{-0pt}{$s$}}
\vc{\includegraphics{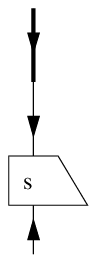}}
\hspace{-10pt}
}
\]
Applying complex conjugation to $p ^\dag \circ p = \id _{A}$ we obtain $p ^* \circ p _* = \id _{A ^*}$, and applying this to the equation pictured above we obtain $s = p ^* \circ t \circ p$. Also, from the monoid homomorphism equation $p \circ u = v$ we obtain $u = p ^\dag \circ v$, and therefore $u ^\dag = v ^\dag \circ p$ by applying the \dag\-functor. Using these equations, along with the multiplication compatibility equation $p \circ m = n \circ (p \otimes p)$, we obtain the following:
\[
{
\lessa
\psfrag{p}{$s$}
\psfrag{ps}{$\,\,\,\,p^*$}
\vc{\includegraphics{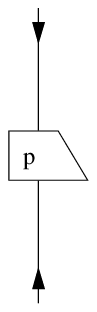}}
\lessa
}
=
{
\lessa
\psfrag{p}{$p$}
\psfrag{ps}{$\,\,\,\,p^*$}
\vc{\includegraphics{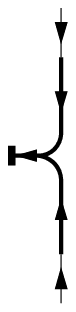}}
\lessa
}
=
{
\lessa
\psfrag{s}{$p$}
\vc{\includegraphics{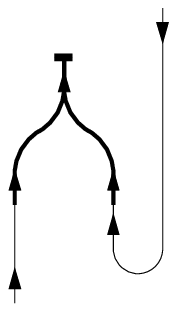}}
\lessa
}
=
{
\lessa
\psfrag{p}{$p$}
\vc{\includegraphics{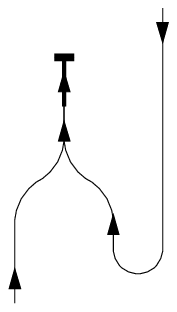}}
\lessa
}
=
{
\lessa
\psfrag{s}{$p$}
\vc{\includegraphics{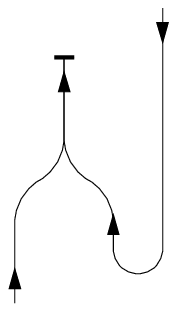}}
\lessa
}
=
{
\lessa
\psfrag{s}{$p$}
\vc{\includegraphics{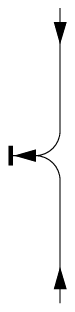}}
\lessa
}
\]
The involution is therefore the left involution associated to the monoid.

We now show that the monoid is in fact a \dag\-Frobenius monoid. To start with we use the fact that $p$ is an isometry and that it preserves multiplication, along with the unit law of the monoid and the Frobenius law:
\[
{
\lessa
\vc{\includegraphics{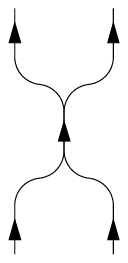}}
\lessa
}
=
{
\lessa
\vc{\includegraphics{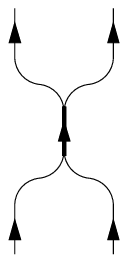}}
\lessa
}
=
{
\lessa
\vc{\includegraphics{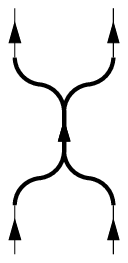}}
\lessa
}
=
{
\lessa
\vc{\includegraphics{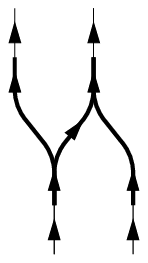}}
\lessa
}
=
{
\lessa
\vc{\includegraphics{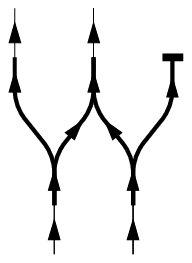}}
\lessa
}
=
{
\lessa
\vc{\includegraphics{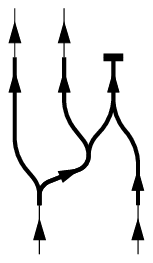}}
\lessa
}
\]
We now employ the fact that $p$ preserves the involution, and then essentially perform the previous few steps in reverse order:
\[
{
\lessa
\vc{\includegraphics{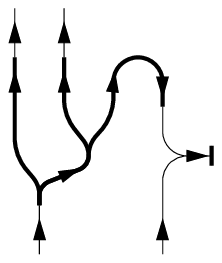}}
\lessa
}
=
{
\lessa
\vc{\includegraphics{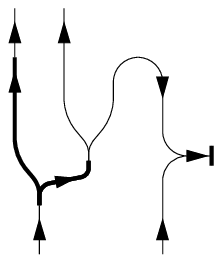}}
\lessa
}
=
{
\lessa
\vc{\includegraphics{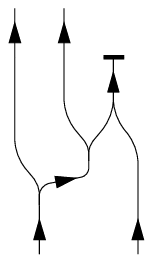}}
\lessa
}
=
{
\lessa
\vc{\includegraphics{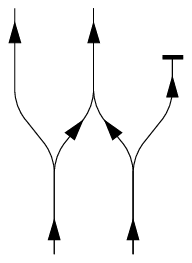}}
\lessa
}
=
{
\lessa
\vc{\includegraphics{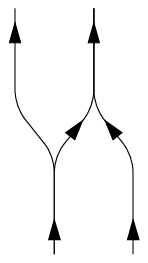}}
\lessa
}
\]
The proof for the other Frobenius law is exactly analogous. We have demonstrated that the monoid $(A,m,u)$ is \dag\-Frobenius, and since we have shown that the involution $s$ is the left involution associated to the monoid, it follows that $(A,m,u; s)$\ is a \dag\-Frobenius left-involution monoid.
\end{proof}

\section{Special unitary \dag\-Frobenius monoids in \cat{Hilb}}
\label{hilbsect}

From now on we will mainly work in \cat{Hilb}, the category of finite-dimensional complex Hilbert spaces and linear maps, which is a symmetric monoidal \dag\-category with duals. Special unitary \dag\-Frobenius monoids have particularly good properties in this setting.

The following lemma contains the important insight due to Coecke, Pavlovic and the author, as described in the introduction and in \cite{cpv08-dfb}.

\begin{lemma}
\label{smallcalglemma}
In \cat{Hilb}, a \dag\-Frobenius right-involution monoid admits a norm making it into a C*\-algebra.
\end{lemma}
\jbeginproof
By Lemma~\ref{involembedding} a \dag\-Frobenius right-involution monoid $(A,m,u)$ has an involution-preserving embedding into $\End(A)$, which is a C*\-algebra when equipped with the operator norm. The involution monoid $(A,m,u)$ therefore admits a C*\-algebra norm, taken from the norm on $\End(A)$ under the embedding. Since the algebra is finite-dimensional, the completeness requirement is trivial.
\end{proof}

We will also require the following important result, which demonstrates a crucial abstract property of the category \cat{Hilb}.

\begin{lemma}
\label{hilbisocounit}
In \cat{Hilb}, isomorphisms of special unitary \dag\-Frobenius involution monoids preserve the counit.
\end{lemma}
\jbeginproof
Any special unitary \dag\-Frobenius involution monoid is in particular a \dag\-Frobenius right-involution monoid, and so admits a norm with which it becomes a C*\-algebra by Lemma~\ref{smallcalglemma}. Finite-dimensional C*\-algebras are semisimple, and are therefore isomorphic to finite direct sums of matrix algebras in a canonical way; an isomorphism between two finite-dimensional C*\-algebras is then given by a direct sum of pairwise isomorphisms of matrix algebras. We therefore need only show that the lemma is true for special unitary \dag\-Frobenius involution monoids which are matrix algebras, with involution given by matrix adjoint.

Let $(A,m,u;s)$ and $(B,n,v;t)$ be special unitary \dag\-Frobenius involution monoids which are both isomorphic to some matrix algebra $\End(\mathbb{C} ^n)$. Any isomorphism between them must have some decomposition into isomorphisms $f:(A,m,u;s) \to \End(\mathbb{C} ^n)$ and \linebreak\mbox{$g: \End(\mathbb{C} ^n) \to (B,n,v;t)$}. The statement that $g \circ f$ preserves the counit is equivalent to the statement that the outside diamond of the following diagram commutes:
\begin{equation}
\begin{diagram}[midshaft,height=20pt,width=30pt,nohug]
&& \mathbb{C} &&
\\
&
\ruTo(2,2) ^{u ^\dag}
&&
\luTo(2,2) ^{v ^\dag}
&
\\
(A,m,u;s) && \uTo>{\mathrm{Tr}} && (B,n,v;t)
\\
&
\rdTo(2,2) _f ^\simeq
&&
\ruTo(2,2) _g ^\simeq
&
\\
&& \End(\mathbb{C} ^n) &&
\end{diagram}
\end{equation}

We will show that each triangle separately commutes, and therefore that the entire diagram commutes. We focus on the triangle involving the isomorphism $g$; the treatment of the other triangle is analogous. Our strategy is to show that $\rho_g :=\nfrac \cdot v ^\dag \circ g$ is a tracial state of $\End(\mathbb{C} ^n)$. It takes the unit to 1, since $\nfrac \cdot v ^\dag \circ g \circ \epsilon _B ^\pL = \nfrac \cdot v ^\dag \circ v = \nfrac \cdot \dim(B) = \nfrac \cdot n = 1$, where we used the fact that $g$ is a homomorphism and Lemma~\ref{dimnormu}; this is the reason that we require the \dag\-Frobenius monoid to be special. We can simplify the action of $\rho _g$ on positive elements in the following way, where $\phi : I \to \mathbb{C} ^n {}^* \otimes \mathbb{C}^n$ is an arbitrary nonzero state of $\End(\mathbb{C} ^n)$, and $\phi'$ is the result of applying the involution to this state:
\begin{gather*}
\psfrag{r}{$\!n\rho_g$}
\psfrag{p}{$\,g$}
\psfrag{f}{\raisebox{-1.5pt}{$\,\phi$}}
\psfrag{fp}{\raisebox{-1.5pt}{$\,\phi'$}}
\psfrag{g}{$\,g$}
\psfrag{fd}{\raisebox{-1.9pt}{$\,\phi ^\dag$}}
\psfrag{gd}{\raisebox{-1.9pt}{\,$g ^\dag$}}
\psfrag{n}{}
\vc{\includegraphics[scale=0.9]{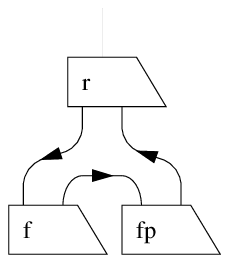}}
\hspace{-15pt}
=
\hspace{-10pt}
\vc{\includegraphics[scale=0.9]{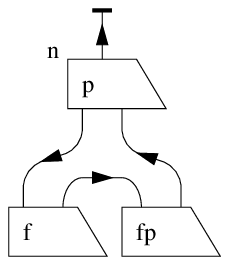}}
\hspace{-15pt}
=
\hspace{-5pt}
\vc{\includegraphics[scale=0.9]{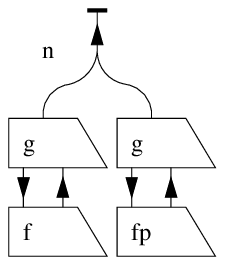}}
\hspace{-10pt}
=
\hspace{-5pt}
\vc{\includegraphics[scale=0.9]{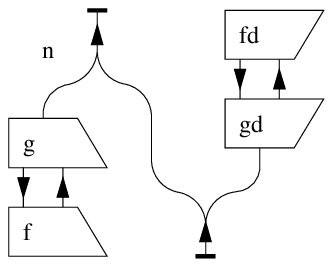}}
\hspace{-10pt}
=
\hspace{-15pt}
\vc{\includegraphics[scale=0.9]{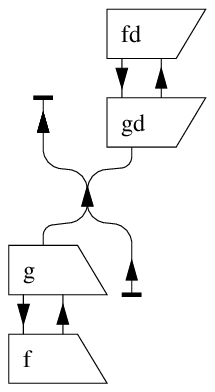}}
\hspace{-15pt}
=
\hspace{-10pt}
\vc{\includegraphics[scale=0.9]{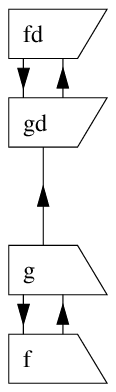}}
\end{gather*}

\jnoindent
The expression on the right-hand side is the squared norm of $g \circ \phi$, which is positive because the inner product in \cat{Hilb} is nondegenerate and $\phi$ is nonzero; this shows that $\rho _g$ takes positive elements to nonnegative real numbers, and so is a state of $\End( \mathbb{C} ^n)$. By Lemma~\ref{unitarylemma} the involution monoid $\End(A)$ is balanced-symmetric, and since we are in \cat{Hilb}, the balancing loop can be neglected; this means that $\rho _g \circ (a \otimes b) = \rho _g \circ (b \otimes a)$ for all $a,b \in \End(A)$, and so $\rho_g$ is tracial.
Altogether $\rho_g$ is a tracial state of a matrix algebra. However, it is a standard result that the matrix algebra on a complex $n$-dimensional vector space has a unique tracial state given by $\nfrac \mathrm{Tr}$ (for example, see \cite[Example~6.2.1]{m90-caot}). It follows that $\rho_g = \frac{1}{n} \mathrm{Tr}$, and so the triangle commutes as required.
\end{proof}

\jnoindent
We can combine this with an earlier lemma to obtain a very useful result.
\begin{lemma}
\label{isosareunitary}
In \cat{Hilb}, isomorphisms of special unitary \dag\-Frobenius involution monoids are unitary.
\end{lemma}

\jbeginproof
Straightforward from Lemmas~\ref{counitunitary} and \ref{hilbisocounit}.
\end{proof}

Given a \dag\-Frobenius monoid in \cat{Hilb}, we will show that scaling the inner product on the underlying complex vector space produces a family of new \dag\-Frobenius monoids. We first note the following relationship between scaling inner products and adjoints to linear maps.

\begin{lemma}
\label{scalelemma1}
\begin{sloppypar}
Let $V$ be a complex vector space with inner product $(-,-) _V$ and let \mbox{$f:V ^{\otimes n} \to V ^{\otimes m}$} a linear map, with the adjoint $f ^\dag$ under this inner product. If the inner product is scaled to $\alpha \cdot (-,-) _V$ for $\alpha$ a positive real number, the adjoint to $f$ becomes $\alpha^{m-n} f ^\dag$.
\end{sloppypar}
\end{lemma}
\jbeginproof
\begin{sloppypar}
Writing the scaled inner product as $(\!( -,-)\!) _V$ and denoting the adjoint to $f$ under this scaled inner product as $f ^\ddagger$, we must have \mbox{$(\!(f\circ x,y) \!) _{V ^{\otimes m}} = ( \!( x, f ^\ddagger \circ y ) \! ) _{V ^{\otimes n}}$}. Using $(\!(-,-)\!) _{V ^{\otimes n}} = \alpha ^n \cdot(-,-) _{V ^{\otimes n}}$ and making the substitution \mbox{$f ^\ddagger = \alpha  ^{m-n} f ^\dag$}, we obtain \mbox{$(f \circ x, y) _{V ^{\otimes m}} = (x, f ^\dag \circ y) _{V ^{\otimes n}}$} which holds by the definition of $f ^\dag$, and so $f ^\ddagger$ is a valid adjoint to $f$ under the new inner product.
\qedhere
\end{sloppypar}
\end{proof}

\begin{lemma}
\label{scalelemma2}
For a \dag\-Frobenius monoid $(A,m,u)$, scaling the inner product on $A$ by any positive real number gives rise to a new \dag\-Frobenius monoid. Moreover, this scaling preserves unitarity.
\end{lemma}
\jbeginproof
This is easy to show using the previous lemma. The \dag\-Frobenius equations will all be scaled by the same factor since they are all composed from a single $m$ and $m ^\dag$, so they will still hold. The unitarity property is an equation involving an $m$ and a $u ^\dag$ on each side, and so both sides of this equation will also scale by the same factor.
\end{proof}

We are now ready to prove our main correspondence theorem between finite-dimensional C*\-algebras and symmetric unitary \dag\-Frobenius monoids.
\begin{theorem}
\label{maincalgthm}
In \cat{Hilb}, the following properties of an involution monoid are equivalent:
\jbeginenumerate
\item it admits a norm making it a C*\-algebra;
\item it admits an inner product making it a special \mbox{unitary \dag\-Frobenius involution monoid};
\item it admits an inner product making it a \dag\-Frobenius right-involution monoid.
\end{enumerate}
\jnoindent
Furthermore, if these properties hold, then the structures in 1 and 2 are admitted uniquely.
\end{theorem}

\jbeginproof
First, we point out that the norm of property 1 is \emph{not} directly related to the inner products of properties 2 or 3, in the usual way by which a norm can be obtained from an inner product, and sometimes vice-versa. In fact, the norm of a C*\-algebra will usually not satisfy the parallelogram identity, and so cannot arise directly from any inner product.

We begin by showing $1 \Rightarrow 2$. We first decompose our finite-dimensional C*\-algebra into a finite direct sum of matrix algebras. For any such matrix algebra, an inner product is given by $(a,b) := \mathrm{Tr}(a ^\dag b)$, which is normalized such that $\mathrm{Tr}(\id)=n$ for a matrix algebra acting on $\mathbb{C} ^n$. This gives an endomorphism monoid $\End(\mathbb{C} ^n)$ in \cat{Hilb} for each $n$, which is a unitary \dag\-Frobenius monoid as described by Lemmas \ref{endoisfrob} and \ref{endoequalinv}. Such a monoid is not special unless it is one-dimensional; we have $m \circ m ^\dag = n \cdot \id _{A^* \otimes A}$, where $m$ is the multiplication for the endomorphism monoid. We rescale the inner product, replacing it with $(\!(a,b)\!) := n\, \mathrm{Tr}(a ^\dag b)$. As described by Lemma~\ref{scalelemma1}, writing the adjoint of $m$ under this new inner product as $m ^\ddagger$, we will have $m ^\ddagger = \nfrac m ^\dag$, and $m \circ m ^\ddagger = \id _{A ^* \otimes A}$. By Lemma~\ref{scalelemma2} this preserves the involution and the unitarity of the monoid, and so we obtain a special unitary \dag\-Frobenius monoid with the same underlying algebra and involution as the original matrix algebra. Taking the direct sum of these for each matrix algebra in the decomposition gives a special unitary \dag\-Frobenius involution monoid, with the same underlying algebra and involution as the original C*\-algebra.

The implication $2 \Rightarrow 3$ is trivial, and the implication $3 \Rightarrow 1$ is contained in Lemma~\ref{smallcalglemma}, so the three properties are therefore equivalent.

We now show that, if these properties hold, the norm and inner product in properties 1 and 2 are admitted uniquely. It is well-known that a C*\-algebra admits a unique norm. Now assume that a finite-dimensional complex \mbox{$*$-algebra} has two distinct inner products, which give rise to two special unitary \dag\-Frobenius involution monoids. Since these monoids have the same underlying set of elements and the same involution, there is an obvious involution-preserving isomorphism between them given by the identity on the set of elements. But by Lemma~\ref{isosareunitary} any isomorphism of special unitary \mbox{\dag\-Frobenius} involution monoids in \cat{Hilb} is necessarily an isometry, and therefore unitary, and so the inner products on the two monoids are in fact the same.
\end{proof}

As a result, we can demonstrate some equalities and equivalences of categories.

\begin{theorem}
The category of finite-dimensional C*\-algebras is
\jbeginenumerate
\item
equal to the category of special unitary \dag\-Frobenius involution monoids in \cat{Hilb};
\item
equivalent to the category of unitary \dag\-Frobenius involution monoids in \cat{Hilb}; and
\item
equivalent to the category of \dag\-Frobenius right-involution monoids in \cat{Hilb};
\end{enumerate}
\jnoindent
where all of these categories have involution-preserving monoid homomorphisms as morphisms.
\end{theorem}
\jbeginproof
We prove 1  by noting that the objects of the category of finite-dimensional \mbox{C*\-algebras} are the same as the objects in the category of special unitary \dag\-Frobenius involution monoids in \cat{Hilb}, since in both cases they are involution monoids satisfying one of the first two equivalent properties of Theorem~\ref{maincalgthm}, which can only be satisfied uniquely. The morphisms are also the same, and so the categories are equal.

For 2 and 3, we note that both of these types of structure admit C*\-algebra norms by Lemma~\ref{smallcalglemma}. This gives rise to functors from the categories of 2 and 3 to the category of finite-dimensional C*\-algebras. These functors are full and faithful on hom-sets, since the hom-sets have precisely the same definition in both categories, consisting of all involution-preserving algebra homomorphisms. These functors are also surjective on objects, since given a finite-dimensional C*\-algebra, by Theorem~\ref{maincalgthm} we can find an inner product on the underlying vector space such that the $*$-algebra is in fact a special unitary \dag\-Frobenius involution monoid. Recall that the latter are the objects in the categories of 2 and 3. Since the two functors are full, faithful and surjective, they are therefore equivalences.
\end{proof}

\jnoindent
Our use of the adjective `equal' here perhaps deserves some explanation. It is only appropriate given the way that we have defined the categories of C*\-algebras and of special unitary \dag\-Frobenius monoids, with objects being $*$-algebras that have the \emph{property} of admitting an appropriate norm or inner product. Had we instead defined the objects as being $*$-algebras \emph{equipped} with their norm or inner product, then the categories would not be equal but isomorphic.

\begin{sloppypar}
Having demonstrated the equivalence between finite-dimensional C*\-algebras and \mbox{\dag\-Frobenius} monoids, it becomes clear that Lemmas \ref{embedmonoidlemma} and \ref{involembedding} are precisely the finite-dimensional noncommutative Gelfand-Naimark theorem, that any abstract finite-dimensional C*\-algebra has an involution-preserving embedding into the algebra of bounded linear operators on a Hilbert space. It is striking that these lemmas are quite easy to prove from the \dag\-Frobenius monoid point of view, compared to the traditional \mbox{C*\-algebra} perspective. However, to prove Theorem~\ref{maincalgthm} we used the decomposition theorem for finite-dimensional C*\-algebras from which the finite-dimensional noncommutative Gelfand-Naimark theorem trivially follows, so this does not constitute a new proof; for this, we would need a more direct way to establish the link between finite-dimensional C*\-algebras and \dag\-Frobenius monoids.
\end{sloppypar}

In contrast, some properties of C*\-algebras are \emph{harder} to demonstrate from the perspective of \dag\-Frobenius monoids, as demonstrated by Lemma~\ref{involembedding}. The proof of that lemma required 14 applications of identities, while the corresponding property of finite-dimensional C*\-algebras, that  any involution-closed subalgebra is also a C*\-algebra, is trivial.

\section{Generalizing the spectral theorem}
\label{genspec}
\subsubsection*{Classical structures and spectral categories}

As a consequence of being able to define finite-dimensional C*\-algebras internally to a category, we are also able to state the finite-dimensional spectral theorem categorically. As an introduction to this, we first give a brief summary of some of the main ideas of \cite{cpv08-dfb}.

We start by introducing an important connection between \emph{commutative} \dag\-Frobenius monoids and finite sets.
\begin{defn}
In a braided monoidal category, a monoid is \emph{commutative} if the braiding and the multiplication satisfy the \emph{commutativity} equation:
\begin{equation}
\vc{\includegraphics{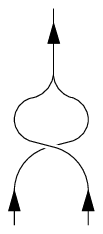}}
=
\vc{\includegraphics{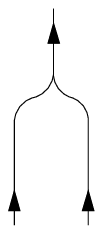}}
\end{equation}
\end{defn}

\begin{theorem}
\label{theoremhilbset}
The category of commutative \dag\-Frobenius monoids in \cat{Hilb} with involution-preserving\footnote{In fact, this involution-preservation condition is not required: as demonstrated in \cite{cpv08-dfb}, every homomorphism of finite-dimensional commutative C*\-algebras is involution-preserving.
} monoid homomorphisms as morphisms is equivalent to the opposite of \cat{FinSet}, the category of finite sets.
\end{theorem}
\jbeginproof
A commutative \dag\-Frobenius monoid in \cat{Hilb} is balanced-symmetric, since the balancing is the identity in that category, and is therefore unitary by Lemma~\ref{unitarylemma}. By Theorem~\ref{maincalgthm}, the category being constructed is therefore isomorphic to the category of finite-dimensional commutative C*\-algebras with algebra homomorphisms as morphisms. We apply the spectral theorem for commutative C*\-algebras to obtain the desired result.
\end{proof}

\jnoindent
Put more straightforwardly, a choice of commutative \dag\-Frobenius monoid on a Hilbert space defines a basis for that Hilbert space. In fact, the bases for each space are in precise correspondence to the \emph{special} commutative \dag\-Frobenius monoids, as might be expected from our Theorem~\ref{maincalgthm}; the same basis will be determined by many different \dag\-Frobenius monoids.

Theorem~\ref{theoremhilbset} motivates the following definition:
\begin{defn}
\label{defclasstruct}
In a braided monoidal \dag\-category, a \emph{classical structure} is a commutative \dag\-Frobenius monoid. If the underlying object is $A$, then we say that it is a \emph{classical structure on $A$}.
\end{defn}

\jnoindent
Classical structures were first described by Coecke and Pavlovic in \cite{cp06-qmws}, and the philosophy of that paper --- that a classical structure represents the possible outcomes of \mbox{a measurement ---} is embraced here.
\begin{defn}
\label{defclasscat}
Given a braided monoidal \dag\-category \cat{Q}, its \emph{category of classical structures} $\C(\cat{Q})$ is the category with classical structures in \cat{Q} for objects, and involution-preserving monoid homomorphisms as morphisms.
\end{defn}

\jnoindent
Using this notation, the result in Theorem~\ref{theoremhilbset} can be written as
\begin{equation}
\label{chilbfinset}
\C(\cat{Hilb}) \simeq \cat{FinSet} ^{\mathrm{op}}.
\end{equation}

\begin{sloppypar}
These results give a new perspective on the relationship between finite-dimensional Hilbert spaces and finite sets. We can construct a covariant forgetful functor \mbox{$\textrm{\emph{Forget}} : \C(\cat{Hilb}) \to \cat{Hilb}$} which takes a classical structure to its underlying Hilbert space. We can also construct a covariant functor \mbox{$\textrm{\emph{Free}} : \cat{FinSet} ^{ \mathrm {op}} \to \cat{Hilb}$}, which takes a set to a Hilbert space freely generated by taking that set as an orthonormal basis, and a function between sets to the adjoint of the linear map that has the same action on the chosen basis. Using the equivalence $\C(\cat{Hilb}) \simeq \cat{FinSet} ^{\mathrm{op}}$ implied by Theorem~\ref{theoremhilbset}, we see that the functors \emph{Forget} and \emph{Free} are naturally isomorphic. We have two quite different points of view, which are both equally valid: a set is a Hilbert space with the extra structure of a special commutative \dag\-Frobenius monoid, and a Hilbert space is a set with the extra structure of a complex vector space.
\end{sloppypar}

One possible point of view is that a classical structure represents a \emph{measurement} performed on the underlying Hilbert space, or rather, on the physical system which has that Hilbert space as its space of states. To say `the possible results of a measurement form a finite set' can then be directly interpreted by the formalism: if we are doing our quantum theory in a braided monoidal \dag\-category \cat{Q}, it is simply the statement that $\C(\cat{Q}) \simeq \cat{FinSet}$. The emergent `classical logic' with which we reason about these measurement results is then more `powerful' when the category $\C(\cat{Q})$ has more interesting properties; for example, it could be a fully-fledged elementary topos, as for the case of \cat{Hilb}. With this in mind, we make the following definition:
\begin{defn}
A braided monoidal \dag\-category \cat{Q} is \emph{spectral} if $\C(\cat{Q})$ is an elementary topos.
\end{defn}
\jnoindent
Spectral categories can be thought of as generalized settings for quantum theory which admit a particularly good `generalized spectral theorem', or in which measurement outcomes admit a particularly good logic. We describe a class of spectral categories in Theorem~\ref{fqBt}, which have finite Boolean topoi as their categories of classical objects.

We briefly mention a connection to other work. \label{didiscussion} D\"oring and Isham \cite{di08-nsp} have developed a topos-theoretic approach to analyzing the logical structure of theories of physics, in which a quantum system is explored through the presheaves on the partially-ordered set of commutative subalgebras of a von Neumann algebra. In finite dimensions von Neumann algebras coincide with C*\-algebras, and therefore also with special unitary \dag\-Frobenius monoids in \cat{Hilb} by Theorem~\ref{maincalgthm}. Given a \dag\-Frobenius monoid of this type, the partially-ordered set of special commutative sub-\dag\-Frobenius monoids can be constructed categorically, and so D\"oring-Isham toposes can be constructed directly  from any special unitary \dag\-Frobenius monoid in any braided monoidal \dag\-category. The techniques of that research program can then be employed; in particular, we can test whether a generalized Kochen-Specker theorem holds. In fact, we suggest that this approach could be used quite generally to connect the ideas of D\"oring and Isham to other work on monoidal categories in the foundations of quantum physics, such as that of Abramsky, Coecke and others~\cite{ac08-cqm, ce08-tqc}.

\subsubsection*{The spectral theorem for normal operators}

We now turn to the spectral theorem for normal operators, which says that a normal operator on a complex Hilbert space can be diagonalized. For complex Hilbert spaces this follows from the spectral theorem for commutative \mbox{C*\-algebras}, since any normal operator generates a commutative \mbox{C*\-algebra} and the spectrum of this algebra performs the diagonalization. This will not necessarily be the case in an arbitrary monoidal \dag\-category, with C*\-algebras replaced by special unitary \dag\-Frobenius monoids. However, we can nonetheless give a direct categorical description of diagonalization.

We proceed by introducing two different categorical properties which capture the geometrical essence of the spectral theorem for normal operators, and then showing that they are equivalent.

\begin{defn}
\label{compatdefn}
In a monoidal category, an endomorphism $f:A \to A$ is \emph{compatible} with a monoid $(A,m,u)$ if the following equations hold:
\begin{equation}
\begin{array}{ccccc}
\psfrag{s}{\raisebox{-0.50pt}{$f$}}
\hspace{-10pt}
\vc{\includegraphics{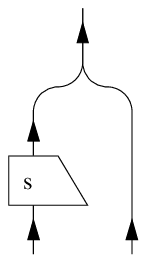}}
\hspace{-10pt}
&=&
\hspace{-20pt}
\psfrag{s}{\raisebox{-0.50pt}{$f$}}
\vc{\includegraphics{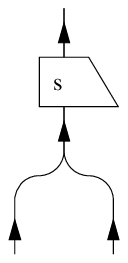}}
\hspace{-20pt}
&=&
\hspace{-10pt}
\psfrag{s}{\raisebox{-0.50pt}{$f$}}
\vc{\includegraphics{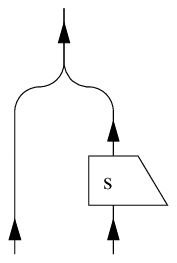}}
\hspace{-10pt}
\hspace{-10pt}
\\
m \circ (f \otimes \id _{A})
&=&
f \circ m
&=&
m \circ (\id _{A} \otimes f)
\end{array}
\end{equation}
\end{defn}

\begin{defn}
\label{intdiag}
In a braided monoidal \dag\-category, an endomorphism $f:A \to A$  is \emph{internally diagonalizable} if it can be written as an action of an element of a commutative \dag\-Frobenius algebra on $A$; that is, if it can be written as
\begin{equation}
\begin{array}{ccc}
\hspace{-10pt}
\psfrag{s}{\raisebox{-0.50pt}{$f$}}
\vc{\includegraphics{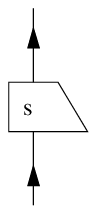}}
\hspace{-10pt}
&
=
&
\hspace{-10pt}
\psfrag{p}{\hspace{-6pt}$\phi _f$}
\vc{\includegraphics{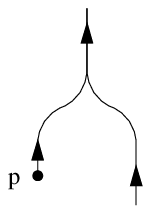}}
\hspace{-10pt}
\\
f
&=&
m \circ (\phi_f \otimes \id_ {A}),
\end{array}
\end{equation}
where $m:A \otimes A \to A$ is the multiplication of a commutative \dag\-Frobenius algebra and \mbox{$\phi_f : I \to A$} is a state of $A$.
\end{defn}

\begin{lemma}
\label{diagcompat}
An endomorphism $f: A \to A$ is internally diagonalizable if and only if it is compatible with a commutative \dag\-Frobenius monoid.
\end{lemma}

\jbeginproof
Assume that $f$ is internally diagonalizable by the action of an element $\phi_f :I \to A$ of a commutative \dag\-Frobenius monoid $(A,m,u)$, so that $f = m \circ (\phi _f \otimes \id _{A})$. The following pictures must be equal by the associativity and commutativity laws, where the multiplication is the morphism $m$:
\[
\psfrag{p}{\hspace{-6pt}$\phi_f$}
{
\hspace{-10pt}
\vc{\includegraphics{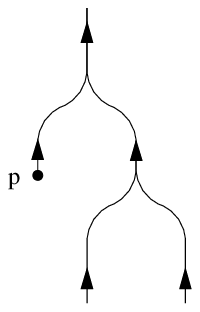}}
\hspace{-10pt}
}
=
{
\hspace{-10pt}
\vc{\includegraphics{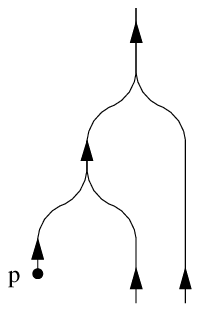}}
\hspace{-10pt}
}
=
{
\hspace{-10pt}
\vc{\includegraphics{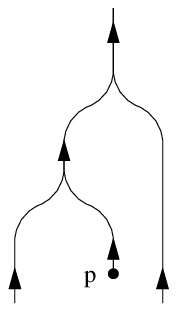}}
\hspace{-10pt}
}
=
{
\psfrag{p}{\hspace{-2pt}$\phi _f$}
\hspace{-10pt}
\vc{\includegraphics{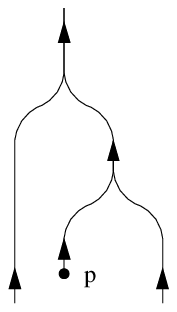}}
\hspace{-10pt}
}
\]
The first picture is $f \circ m$, the second is $m \circ (f \otimes \id _{A})$ and the fourth is $m \circ (\id _{A} \otimes f)$, and so $f$ is compatible with the commutative \dag\-Frobenius monoid $(A,m,u)$. Conversely, assuming compatibility of $f$ with a commutative \dag\-Frobenius monoid $(A,m,u)$ and defining \mbox{$\phi _f = f \circ u$}, we have
\[
m \circ (\phi _f \otimes \id _{A}) = m \circ \jbigl (f \circ u) \otimes \id _{A} \jbigr = f \circ m \circ (u \otimes \id _{A} ) = f
\]
and so $f$ is internally diagonalizable.
\end{proof}

We now show that any internally-diagonalizable endomorphism must be normal, by the properties of commutative \dag\-Frobenius monoids.

\begin{lemma}
\label{diagisnormal}
If an endomorphism $f:A \to A$ is internally diagonalizable, then it is normal.
\end{lemma}
\jbeginproof
The statement that $f$ is internally diagonalizable is equivalent to the statement that $f$ can be written as the left-action of a commutative \dag\-Frobenius monoid. By commutativity this is the same as a right action, and using the notation of the introduction we write this as $R _{\alpha}$ for an element $\alpha \in A$. We then have $f \circ f ^\dag= R _\alpha \circ R _\alpha {}^\dag = R _\alpha \circ R _{\alpha'}$, where $\alpha'$ is defined as in the introduction. By commutativity we have $R _\alpha \circ R _{\alpha'} = R _{\alpha'} \circ R _\alpha$, and so $f \circ f ^\dag = f ^\dag \circ f$.
\end{proof}

\jnoindent
Every internally diagonalizable endomorphism is normal, but is every normal endomorphism internally diagonalizable? This is precisely the content of the conventional spectral theorem for normal operators, and so in \cat{Hilb} the answer is yes.

\begin{lemma}
\label{hilbinternaldiag}
In \cat{Hilb}, every normal endomorphism $f: A \to A$ is internally diagonalizable.
\end{lemma}
\jbeginproof
This follows from the conventional spectral theorem for normal operators. We choose an orthonormal basis set $a_i : \mathbb{C} \to A$, for $1 \leq i \leq \dim(A)$, such that each vector $a_i$ is an eigenvector for $f$. The orthonormal property can be expressed as $a _i ^\dag \circ a_j = \delta _{i j} \id _{\mathbb{C}}$. This basis set is uniquely determined if and only if $f$ is nondegenerate. We use the morphisms $a_i$ to construct a monoid $(A,m,u)$ on $A$ as follows:
\begin{align*}
m &:= \sum _{i=1} ^{\dim(A)} a_i \circ (a_i ^\dag \otimes a_i ^\dag)
\\
u &:= \sum _{i=1} ^{\dim(A)} a_i
\end{align*}
It is straightforward to show that this monoid is in fact a \dag\-Frobenius monoid, which copies the chosen basis for $A$. Since this monoid only copies eigenvectors of $f$ it follows that it is compatible with $f$ in the sense of Definition~\ref{compatdefn}, and so by Lemma~\ref{diagcompat}, the morphism $f$ is internally diagonalizable.
\end{proof}

\subsubsection*{Classical structures in categories of unitary finite-group representations}

\begin{sloppypar}
An important class of `generalizations' of \cat{FinSet} is given by the \textit{finitary toposes}. A \textit{topos}~\cite{m95-ecet} is a category where the operations familiar from traditional constructive logic can all be defined; in particular, unions, products, function sets and powersets are all available. Technically, a topos\footnote{Experts will notice that this is the definition of an \emph{elementary} topos, the most basic type of topos.} is a category with all finite limits, in which every object has a power object; the other constructions just mentioned can then be derived. An example is the category of finite \cat G-sets, for a finite group \cat G: objects are finite sets equipped with a \cat G-action, and morphisms are functions between the underlying sets which are compatible with the group actions. That such a category is in fact a topos is far from obvious, and relies on powerful general theorems \cite{mm92-sgl}.

Given the explicit connection between \cat{FinSet} and \cat{Hilb} established by the equivalence \mbox{$\cat{FinSet} ^\mathrm{op}  \simeq \C(\cat{Hilb})$}, it is natural to ask whether there exist generalizations of \cat{Hilb} which have other finitary topoi as their categories of classical structures. A topos obtained in this way could be interpreted as giving the classical counterpart to a quantum theory, in contrast to the D\"oring-Isham toposes discussed on page~\pageref{didiscussion} which give a direct topos-theoretical view of the quantum structure itself.
\end{sloppypar}

A heuristic argument puts a stumbling block in front of any such attempt.\footnote{I am grateful to Christopher Isham for this argument.} A striking feature of many toposes is that the law of excluded middle can fail, and as a consequence, given a subobject of an object in the topos, the union of the subobject and its complement can fail to give the original object. For a given Hilbert space, a good way to characterize its subobjects is by the projectors on the space. Two projectors $P$ and $Q$ on a Hilbert space represent disjoint subobjects if $PQ = 0$, and in that case their union as subobjects is represented by the projector $P+Q$.

We now work in a category intended as a generalization of \cat{Hilb}, assuming only that it is a \dag\-category with hom-sets which are complex vector spaces. Projectors can be defined in this setting as endomorphisms $P$ satisfying $P ^\dag = P^2 = P$, and we can describe disjointness and union using our categorical structure in the manner just described. Given any projector $P$ we will be able to use the complex vector space structure of the hom-sets to construct a new projector $(1 - P),$ where $1$ is the identity on the space. This new projector is disjoint with $P$, and gives the identity under union with $P$, using the general definitions of these terms given above. In a sense, it therefore seems that the law of excluded middle holds. To avoid this conclusion either the \dag\-functor must go so that projectors cannot be straightforwardly defined, or the complex numbers must go so that we cannot ask that the hom-sets be vector spaces over them, but both are core parts of the mathematical formalism of quantum mechanics which cannot be lightly abandoned.

We will skirt around this argument by focusing on those toposes for which the excluded middle \emph{does} hold: the Boolean toposes, or at least a finitary subclass of these. We will focus on the following types of category:

\begin{defn}
\label{fqBt}
A \emph{finite quantum Boolean topos} is a symmetric monoidal \dag\-category which has a strong symmetric monoidal \dag\-equivalence to a category \cat{Hilb ^G} of finite-dimensional unitary representations of some finite groupoid \cat{G}, where \cat{Hilb} is the category of finite-dimensional complex Hilbert spaces and continuous linear maps.
\end{defn}

\begin{defn}
\label{fBt}
A \emph{finite Boolean topos} is a category equivalent to a topos of the form \cat{FinSet ^G} for some finite groupoid \cat{G}, where \cat{FinSet} is the topos of finite sets and functions.
\end{defn}

\begin{theorem}
\label{frobtopos}
The category of classical structures in a finite quantum Boolean topos is equivalent to a finite Boolean topos, and every finite Boolean topos arises in this way.
\end{theorem}
\jbeginproof
Let \cat{Q} be a finite quantum Boolean topos, for which by definition there exists a strong symmetric monoidal \dag\-equivalence \mbox{$\cat{Q} \simeq \cat{Hilb ^G}$} for a finite groupoid \cat{G}. There is a canonical forgetful \dag\-preserving functor $F: \cat{Hilb ^G} \to \cat{Hilb}$ that takes a unitary \cat{G}\-representation to the Hilbert space on which \cat G is acting. By abuse of notation we will also write $F: \cat{Q} \to \cat{Hilb}$, suppressing the equivalence $\cat{Q} \simeq \cat{Hilb ^G}$. A commutative \dag\-Frobenius monoid $( A, m, u )$ in \cat{Q} gives a commutative \dag\-Frobenius monoid $\jbigl F(A), F(m), F(u) \jbigr$ in \cat{Hilb}, and therefore defines a basis for the Hilbert space $F(A)$ by Theorem~\ref{theoremhilbset}. Each object $A$ of \cat{Q}, via the equivalence with \cat{Hilb ^G}, is actually a \dag\-functor $A :\cat{G} \to \cat{Hilb}$, and for each $g \in \cat{G}$ the morphism \mbox{$A(g) : F(A) \to F(A)$} is a unitary linear map in \cat{Hilb}. The morphisms $F(m)$ and $F(u)$ are intertwiners, which can be expressed by the following commuting diagram that holds for all $g \in \cat{G}$:
\[
\begin{diagram}[midshaft,width=75pt,height=23pt]
F(A) \otimes F(A) & \rTo ^{A (g) \otimes A(g)} & F(A) \otimes F(A)
\\
\dTo <{ F(m) } && \dTo > {F(m)}
\\
F(A) & \rTo ^{A (g)} & F(A)
\\
\uTo <{F(u)} && \uTo >{F(u)}
\\
F(I) & \rEq & F(I)
\end{diagram}
\]

Read differently, this diagram is also precisely the condition for $A(g)$ to be a monoid homomorphism for the commutative \dag\-Frobenius monoid $\jbigl F(A), F(m), F(u) \jbigr$ in \cat{Hilb}. Since the morphism $A(g)$ is invertible, it must act as a permutation of the basis of $F(A)$ defined by the monoid, and the commutative \dag\-Frobenius monoid $( A,m,u )$ therefore corresponds to an action of the groupoid \cat{G} on this basis. Every finite \cat{G}-action must arise in this way, since any \cat{G}-action on a finite set gives rise to a linear \cat{G}-representation on the complex Hilbert space with basis given by elements of the set. Morphisms between commutative \mbox{\dag\-Frobenius} monoids have adjoints which act as set-functions for the induced bases, and these adjoints are compatible with the induced \cat{G}-actions on the basis elements. It follows that the category of commutative \dag\-Frobenius monoids in \cat{Q \simeq Hilb ^G} is equivalent to the opposite of the category \cat{FinSet ^G}.
\end{proof}

Another way to phrase this result is that the process of taking \cat{G}-presheaves --- either of sets, or of Hilbert spaces --- commutes with the process of forming the category of classical objects:
\begin{equation}
\C(\cat{Hilb} ^{\cat{G}}) \simeq \C(\cat{Hilb}) ^{\cat{G}} \simeq \cat{FinSet^G}.
\end{equation}

\jnoindent
For the functor category \cat{Hilb ^G} we take only unitary representations, or equivalently \mbox{\dag\-preserving} functors where the \dag\-functor on \cat{G} takes a morphism to its inverse. It is this result which motivates the term `finite quantum Boolean topos'. We also note that we can use this to recover the finite groupoid \cat{G} from its unitary representation category \cat{Hilb^G}, since \cat{FinSet^G} yields \cat{G} as its smallest full generating subcategory (see \cite[Chapter~6]{mm92-sgl}). 

Given the similarity between the presheaf-style definitions  \ref{fqBt} and \ref{fBt}, the lemma perhaps seems artificial. In fact, it is known that finite quantum Boolean toposes can be described axiomatically; it follows from the Doplicher-Roberts theorem~\cite{dr89-ndt} that, using the terminology of Baez \cite{b97-hda2}, they are precisely the finite-dimensional even symmetric \mbox{2-H*-algebras}. We also expect that finite Boolean toposes would admit a direct axiomatization, although we do not attempt to give one here.

Given the result described here it is interesting to consider a generalization to arbitrary finite-dimensional symmetric 2-H*-algebras. By a generalization of the Doplicher-Roberts theorem \cite{b97-hda2, hm06-aqft} these are known to be the representation categories of finite \emph{supergroupoids}. However, we are not aware of any extensions of our results that can be proved along these lines.

\end{document}